\RequirePackage{fix-cm}
\documentclass[smallextended]{svjour3}       % onecolumn (second format)
\smartqed  % flush right qed marks, e.g. at end of proof
\usepackage{graphicx}
\usepackage{dcolumn}% Align table columns on decimal point
\usepackage{bm}% bold math
\usepackage{amssymb}
\usepackage{amsmath}
\usepackage{amsfonts,mathrsfs}

\begin{document}

\title{ A time-symmetric soliton dynamics \`a la de Broglie}

%\titlerunning{On the existence of Lorentz-invariant}        % if too long for running head

\author{Aur\'elien Drezet}


\institute{A. Drezet \at Institut NEEL, CNRS and Universit\'e Grenoble Alpes, F-38000 Grenoble, France \\
            \email{aurelien.drezet@neel.cnrs.fr} }

\date{Received: date / Accepted: date}

\maketitle

\begin{abstract}
In this work we develop a time-symmetric soliton theory for quantum particles inspired from works by de Broglie and Bohm.  We consider explicitly a non-linear Klein-Gordon theory leading to monopolar oscillating solitons.   We show that the theory is able to reproduce the main results  of the pilot-wave interpretation for non interacting  particles in a external electromagnetic field.  In this regime, using  the time symmetry of the theory,  we are also able to explain quantum entanglement between several solitons and we reproduce the famous pilot-wave nonlocality associated with the de Broglie-Bohm theory.      
\keywords{De Broglie double solution \and Soliton \and Time symmetry \and Bohmian mechanics}
\end{abstract}

\section{Introduction}\label{sec1}
\indent Seventy years ago David Bohm~\cite{Bohm1952}, rediscovering some older results made by de Broglie \cite{Valentini,deBroglie1927},  published his deterministic hidden-variables theory showing that quantum mechanics can be reproduced by a dynamics where particles follow  trajectories guided by a $\psi-$wave solution of the Schr\"odinger equation. This pilot wave theory (PWI) is clearly a counter example against the complacency of the previous period when even the possible existence of hidden variable was contested (e.g., by the von Neumann theorem).   Moreover,  the PWI is counter-intuitive: It involves a nonlocal `spooky' quantum potential in tension with the theory of relativity, and there is no back-reaction of the particle on the guiding wave (which nature is by the way unclear). For these reasons the proposal made by Bohm is often rejected or criticized.  That was the case of Louis de Broglie who actually invented the PWI already in 1926\cite{Valentini,deBroglie1927}  but favored a different approach namely the double solution program (DSP) where a particle is a kind of singularity or localized wave (i.e., a soliton) in an oscillating field guiding its motion \cite{deBroglie1927,deBroglie1956} (for reviews see \cite{Fargue,Fargue2,Durt,Drezet1}). The DSP was motivated by classical works made by Poincar\'e, Abraham, Mie \cite{Mie1912}, and Einstein to understand particles and waves as merging objects in a deeper (local) field  theory. However, due to the constraints imposed by quantum mechanics and Bell's theorem the DSP was never successfully developed.\\
\indent In the present work we develop such a theory for a scalar $u-$field solution of a non-linear Klein-Gordon (NLKG) equation involving moving solitons in a external electromagnetic field. As we will see by choosing a specific non-linear term in the NLKG equation, and by using `a phase harmony' condition reminiscent of de Broglie DSP, a self-consistent model can be developed where the soliton core is obeying a dynamics recovering the usual PWI. In our approach the core of the soliton is guided by the phase of the linear Klein-Gordon (LKG) equation. In turn, since the model is completely local, we show that retrocausality associated with waves propagating forward and backward in time is needed in order to reproduce the nonlocal properties of the PWI for several entangled solitons. In the model the nonlocality is thus not fundamental but just effective and results from watching the particle trajectories while ignoring the  underlying retrocausal $u-$field that propagates in space-time and guides the solitons. The present work modifies an earlier analysis \cite{Submitted} in which a nonlocal soliton theory was developed in order to recover the PWI. Moreover, many of the mathematical results derived in \cite{Submitted} are still valid and used in the present work. 
%%%%%%%%%ù
\section{The soliton model and its near-field }\label{sec2}
\begin{figure}[h]
\centering
\includegraphics[width=3 cm]{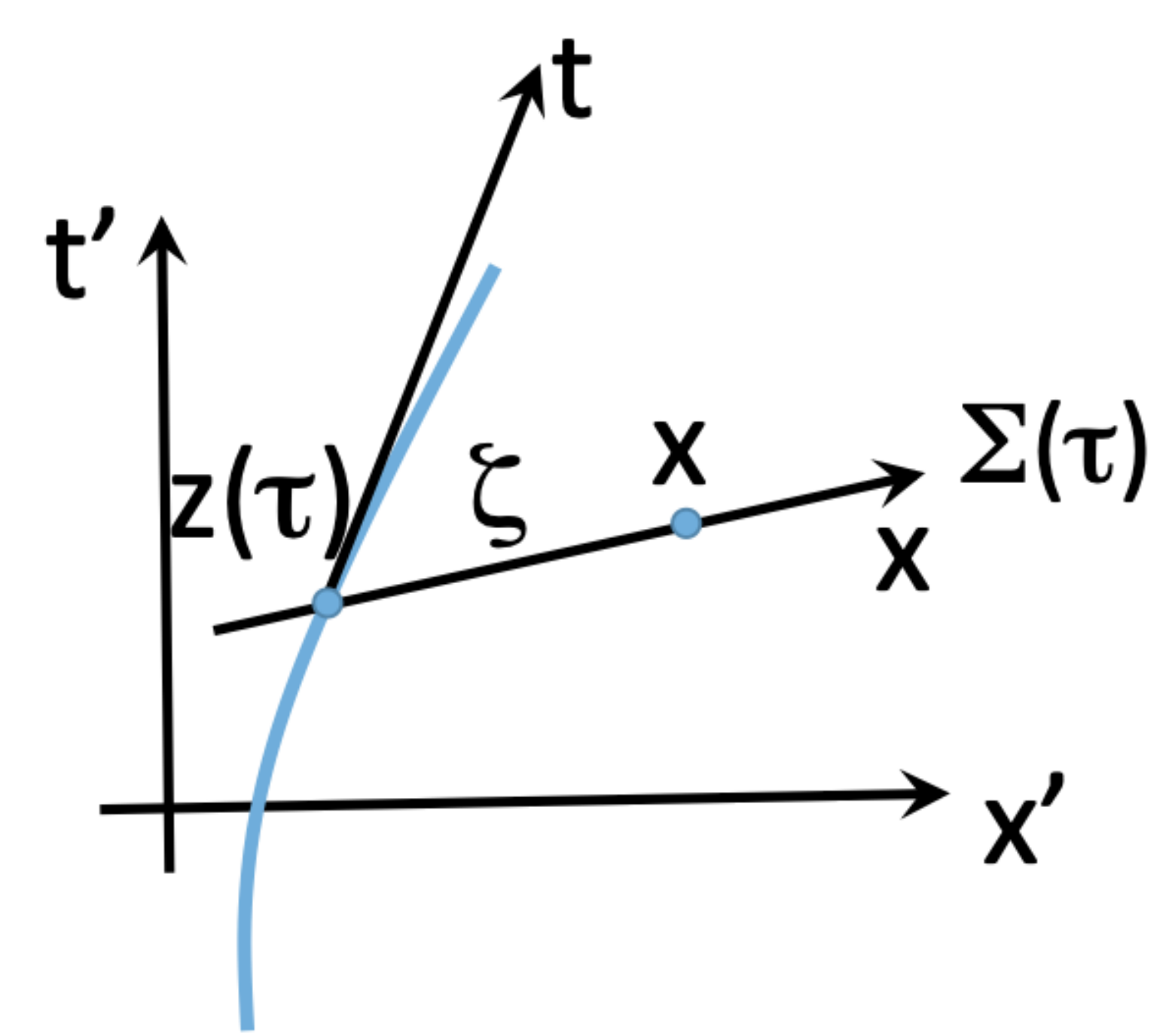}
\caption{The soliton center trajectory $z(\tau)$ (blue curve) seen from the laboratory reference-frame $t',\mathbf{x}'$. $\Sigma(\tau)$ is the hyperplane defined by the velocity $\dot{z}(\tau)$ (i.e., any point $x$ belonging to $\Sigma(\tau)$ is such that $\xi=x-z(\tau)$ is normal to $\dot{z}(\tau)$). The time axis $t$ (defined by $\dot{z}(\tau)$) is tangent to the trajectory at the point $z(\tau)$. In this tangent reference frame $\mathcal{R}_\tau$  we write $x:=[t,\mathbf{x}]$ and a point belonging to the hyperplane $\Sigma(\tau)$ has coordinates $x=[0,\mathbf{x}]$ whereas $z(\tau):=[0,0]$ and thus $\xi:=[0,\mathbf{x}]$.}    \label{image1}
\end{figure}

\indent We start~\cite{Submitted} with the Lagrangian density  $\mathcal{L}=Du(x)D^\ast u^\ast(x)-U(u^\ast (x)u(x))$ for a scalar complex field $u(x)\in\mathbb{C}$ with $x:=[t,\mathbf{x}]\in \mathbb{R}^4$, and with\footnote{We use the  Minkowski metric $\eta_{\mu\nu}$ with signature $+,-,-,-$ and the convention $\hbar=1$, $c=1$.} $D=\partial+ieA(x)$ ($A(x):=[V(x),\mathbf{A}(x)]$ is the electromagnetic potential four-vector and $e$ an electric charge). The nonlinear function $U(y)$ leads to the (Euler-Lagrange) NLKG equation:
\begin{eqnarray}
D^2u(x)=-N(u^\ast (x)u(x))u(x)\label{1b}
\end{eqnarray}   with $N(y):=\frac{dU(y)}{dy}$. This wave equation is different from the linear Klein-Gordon (LKG) equation 
\begin{eqnarray}
D^2\Psi(x)=-\omega_0^2\Psi(x)\label{1c}
\end{eqnarray}  where $ \Psi(x)$ is the standard quantum (relativistic) wavefuntion for a particle of mass $\omega_0$. \\ 
\indent Next, we use the Madelung/de Broglie representation $u(x)=f(x)e^{i\varphi(x)}$, with $f(x), \varphi(x) \in \mathbb{R}$, in Eq.~\ref{1b} and obtain a  pair of coupled hydrodynamic equations:
\begin{subequations}
\label{NL}
\begin{eqnarray}
(\partial \varphi(x)+eA(x))^2=N(f^2(x))+\frac{\Box f(x)}{f(x)}:=\mathcal{M}^2_u(x)\label{2c}\\
\partial[f^2(x)(\partial \varphi(x)+eA(x))]=0.\label{2d}
\end{eqnarray}
\end{subequations} 
Moreover,  for the LKG equation we write similarly 
$\Psi(x)=a(x)e^{iS(x)}$, with $a(x), S(x) \in \mathbb{R}$, yielding the pair of coupled hydrodynamic equations:
\begin{subequations}
\begin{eqnarray}
(\partial S(x)+eA(x))^2=\omega_0^2+Q_\Psi(x):=\mathcal{M}^2_\Psi(x) \label{2}\\
\partial[a^2(x)(\partial S(x)+eA(x))]=0,\label{2b}
\end{eqnarray}
\end{subequations} with $Q_\Psi(x)=\frac{\Box a(x)}{a(x)}$ the so called quantum potential \cite{Bohm1952,deBroglie1956}. Here we consider only the cases $\mathcal{M}^2_\Psi(x)>0$ avoiding tachyonic trajectories.
In the following we show how these equations are solved in the vicinity of the soliton core associated with the localized particle.\\ 
\indent More precisely, assuming that such a soliton exists as a solution of Eq.~\ref{1b} or Eq.~\ref{NL} we  write $z(\tau)$ the trajectory of the soliton center labeled by the proper time $\tau$.  Associated with this particle motion we  define a local Lorentz (proper) rest-frame $\mathcal{R}_\tau$ and an hyperplane $\Sigma(\tau)$ with normal direction given by the velocity $\dot{z}(\tau)$. From geometrical considerations a point $x$ belonging to $\Sigma(\tau)$ satisfies  the constraint \begin{eqnarray}
\xi_\mu\dot{z}^\mu(\tau):=\xi\dot{z}(\tau)=0\label{hyperplane}
\end{eqnarray} with $\xi=x-z(\tau)$ (in $\mathcal{R}_\tau$ we have $\xi:=[0,\boldsymbol{\xi}]$) (see Fig.~\ref{image1}). As we showed in \cite{Submitted} for points $x\in \Sigma(\tau)$ near $z(\tau)$ we can define univocally  the structure of the soliton using the variable $\boldsymbol{\xi}$ in $\mathcal{R}_\tau$ if $1\gg |\textbf{a}|\cdot|\boldsymbol{\xi}|$ where $\textbf{a}$ is the instantaneous acceleration of the center in the rest-frame $\mathcal{R}_\tau$ (i.e., $\xi\ddot{z}\ll 1$). This is interpreted as a condition for defining the notion of quasi-rigidity of the soliton in a relativistic context.\\ 
%%%%%%%%%%%%%%%%
%%%%%%%%
\indent Assuming this condition of rigidity is fulfilled we now introduce the so-called `phase-harmony condition' inspired from de Broglie's DSP~\cite{deBroglie1927,deBroglie1956}:
\begin{quote}
\textit{To every regular solution $\Psi(x)=a(x)e^{iS(x)}$ of Eq.~\ref{1c} corresponds a localized solution $u(x)=f(x)e^{i\varphi(x)}$ of Eq.~\ref{1b} having \underline{locally} the same phase $\varphi(x)\simeq S(x)$, but with an amplitude $f(x)$ involving a generally moving soliton centered on the path $z(\tau)$ and which is representing the particle.}
\end{quote} 
As in \cite{Submitted} (inspired by some earlier non-relativistic results by Petiau~\cite{Petiau1954a,Petiau1954b,Petiau1955} and others \cite{Birula,Rybakov}) we here write  for points $x\in \Sigma(\tau)$ near $z(\tau)$:
\begin{eqnarray}
\varphi(x) \simeq S(z(\tau)) -eA(z(\tau))\xi+B(z(\tau))\frac{\xi^2}{2}+O(\xi^3)\label{phaseharmony}
\end{eqnarray} which defines the phase-harmony condition up to the second-order approximation in power of $\xi$.  The  scalar function $B(z(\tau)):=B(\tau)$ is a collective coordinate required in order to consider the deformation of the soliton with time $\tau$. We mention~\cite{Submitted} that Eq.~\ref{phaseharmony} is not gauge invariant and presupposes the Coulomb-Gauge constraint  $\boldsymbol{\nabla}\cdot \mathbf{A}=0$ in the local rest frame $\mathcal{R}_\tau$. Moreover, the full theory is naturally gauge invariant.\\
\indent With these properties we can extract from Eq.~\ref{NL} several important results derived in \cite{Submitted}. First,  if we define $v_u(x)$, and $v_\Psi(x)$ the velocity of the NLKG and LKG equations respectively
\begin{subequations}
\label{vitesse}
\begin{eqnarray}
v_u(x)=-\frac{\partial \varphi(x)+eA(x)}{\mathcal{M}_u(x)}\label{vitesseu}\\
v_\Psi(x)=-\frac{\partial S(x)+eA(x)}{\mathcal{M}_\Psi(x)}\label{vitessepsi}
\end{eqnarray}  
\end{subequations} we have along $\Sigma(\tau)$\footnote{In \cite{Submitted} we obtained the expression for $\mathcal{M}_u(x)$ (Eq. 66) without expliciting the first order correction $O(\xi)$.  The details of the calculations show that we have $\mathcal{M}_u(x)\simeq \mathcal{M}_\Psi(z(\tau))+\boldsymbol{\xi}\cdot\boldsymbol{\nabla}\mathcal{M}_\Psi(z(\tau))+O(\xi^2)$. A careful analysis shows that we have  $\partial_\mu\mathcal{M}_u(z(\tau))=\partial_\mu\mathcal{M}_\Psi(z(\tau))$. }
\begin{subequations}
\begin{eqnarray}
v_u(x)\simeq v_\Psi(z(\tau)) + O(\xi),\label{vitesseaprox}\\ \mathcal{M}_u(x)\simeq \mathcal{M}_\Psi(z(\tau)) + O(\xi).\label{massaprox}
\end{eqnarray}\end{subequations}
Moreover, by definition we suppose $v_u(z(\tau))=\dot{z}(\tau)$  and therefore we obtain here a guidance condition
 \begin{eqnarray}
v_u(z(\tau))=\dot{z}(\tau)=v_\Psi(z(\tau))\label{guidance}
\end{eqnarray} as postulated  in the original DSP of de Broglie. The phase-harmony condition that we postulate is thus imposing  $\partial \varphi(z(\tau))=\partial S(z(\tau))$. The two phase waves $\varphi$ and $S$ are thus connected along the curve $z(\tau)$. Yet, we emphasize that we don't here impose the second-order matching $\partial_{\mu,\nu}^2\varphi(z(\tau))=\partial_{\mu,\nu}^2S(z(\tau))$ but only a first-order contact \cite{Submitted} meaning that $\partial_{\mu,\nu}^2\varphi(z(\tau))$ and $\partial_{\mu,\nu}^2S(z(\tau))$ are in general different. The relation $\dot{z}(\tau)=v_\Psi(z(\tau))$ is actually the definition given to the particle velocity in the PWI fixing a first order dynamical law. In this PWI we directly deduce a second order dynamical law \cite{Submitted}:
\begin{eqnarray}
\frac{d}{d\tau}[\mathcal{M}_\Psi(z(\tau))\dot{z}^\mu(\tau)]=\partial^\mu [\mathcal{M}_\Psi(z(\tau))]
+eF^{\mu\nu}(z(\tau))\dot{z}_{\nu}(\tau)
 \label{Newton}
\end{eqnarray}  with $F^{\mu\nu}(x)=\partial^\mu A^\nu(x)-\partial^\nu A^\mu(x)$ the Maxwell tensor field at point $x:=z$.
\indent Moreover, we also deduce in the vicinity of  $z(\tau)$:
\begin{eqnarray}
v_u(x)\partial\ln{[f^2(x)]}+\frac{d}{d\tau}\ln{[\mathcal{M}_\Psi(\tau)]}
=\frac{3B(\tau)}{\mathcal{M}_\Psi(\tau)}+O(\xi) \label{Compressapprox}
\end{eqnarray} where we have used the Lagrangian derivative $\frac{d}{d\tau}:=v_\Psi(z)\partial_z$ for the $\Psi-$field along the particle trajectory  $z(\tau)$. From  Eq.~\ref{Compressapprox} we deduce \begin{eqnarray}
\frac{d}{d\tau}\ln{[f^2(z(\tau))\mathcal{M}_\Psi(\tau)]} 
=\frac{3B(\tau)}{\mathcal{M}_\Psi(\tau)}.\label{newd}
\end{eqnarray} To physically interpret Eq.~\ref{newd} it is interesting to note that from Eq.~\ref{2d}  we have $v_u\partial \ln{(f^2\mathcal{M}_u)}:=\frac{d}{d\tau}\ln{(f^2\mathcal{M}_u)}=-\partial v_u$ and that from relativistic hydrodynamics we can define an elementary comoving 3D fluid volume $\delta^3\sigma_0$ (defined in $\mathcal{R}_\tau$) driven by the fluid motion\footnote{We have the fluid conservation: $v_u\partial \ln{(f^2\mathcal{M}_u\delta^3\sigma_0)}:=\frac{d}{d\tau}\ln{(f^2\mathcal{M}_u\delta^3\sigma_0)}=0$.} and such that  $v_u\partial \ln{(\delta^3\sigma_0)}:=\frac{d}{d\tau}\ln{(\delta^3\sigma_0)}=+\partial v_u$. Regrouping all these conditions and using  $\mathcal{M}_\Psi(\tau)=\mathcal{M}_u(\tau)$ we obtain 
\begin{eqnarray}
-\partial v_u(z(\tau))=-\frac{d}{d\tau}\ln{[\delta^3\sigma_0(z(\tau))]}
=\frac{3B(\tau)}{\mathcal{M}_\Psi(\tau)}=\frac{d}{d\tau}\ln{[f^2(z(\tau))\mathcal{M}_\Psi(\tau)]}\label{newdfluid}
\end{eqnarray} 
which shows that a non-vanishing value for $B(\tau)$ involves a compressibility and `deformability' of the soliton droplet.\\
\indent A final important equation can  be derived in the limit $\xi\ddot{z}\ll 1$ where  we have $|\partial_t^2f|\ll|\boldsymbol{\nabla}^2f|$ in the rest-frame $\mathcal{R}_\tau$. We obtain \cite{Submitted} the partial differential equation for the soliton profile for points $x$ belonging to $\Sigma(\tau)$ and localized near $z(\tau)$:
 \begin{eqnarray}
[\omega_0^2+Q_\Psi(z(\tau))]f(x)+\boldsymbol{\nabla}^2f(x)\simeq N(f^2(x))f(x)
\label{ODE}\end{eqnarray} with $\boldsymbol{\nabla}:=\frac{\partial}{\partial\boldsymbol{\xi}}$. Moreover, we suppose the soliton core size $r_0$ to be much smaller than the Compton wavelength  $\omega_0^{-1}$ or even $\mathcal{M}_\Psi^{-1}$ and therefore in the near-field we have
   \begin{eqnarray}
\boldsymbol{\nabla}^2f(x)\simeq N(f^2(x))f(x).
\label{ODEnf}\end{eqnarray} 
\indent In \cite{Submitted} we showed that it is not in general possible to satisfy simultaneously Eq.~\ref{ODEnf} and   Eq.~\ref{newdfluid} because the condition of existence of an underformable localized wave contradicts Eq.~\ref{newdfluid}.   More precisely, assuming a undeformable solution of  Eq.~\ref{ODEnf} we have $f(z(\tau))=f_0=Const.$ $\forall \tau$, i.e., 
 from Eq.~\ref{newdfluid} $B(\tau)=\frac{1}{3}\frac{d}{d\tau}\mathcal{M}_\Psi(\tau)$. But since the soliton is undeformable we have also $-\partial v_u(z(\tau))=-\frac{d}{d\tau}\ln{[\delta^3\sigma_0(z(\tau))]}=0$ and therefore $B(\tau)=0$. This implies $\mathcal{M}_\Psi(\tau)=Const.$ along the trajectory $z(\tau)$ and contradicts the PWI (where the quantum potential is in general not a constant of motion).\\
 \indent Furthermore, in \cite{Submitted} we also showed  from a version of the Ehrenfest theorem adapted to our nonlinear wave equation that the existence of very small soliton with fastly decaying amplitude with the distance $|\boldsymbol{\xi}|$ generally contradicts the PWI unless $\mathcal{M}_\Psi(\tau)=Const.$ along the trajectory $z(\tau)$. In other words,  a strongly localized soliton obeys a classical dynamics  characterized  by a constant mass $\mathcal{M}_\Psi(\tau)=Const.$ and don't reproduce quantum mechanics.\\
 \indent In order to circumvent these objections against the DSP we need here to relax our constraints and we must define a nonlinear function $N(uu^\ast)$ such that the NLKG equation i) admit a soliton deformable along the trajectory $z(\tau)$, and ii) that the soliton is not too strongly localized in order to avoid the conclusion of Ehrenfest theorem.\\
 \indent For this purpose we here consider a power-law non-linearity (also called Lane-Emden non linearity in the context of astrophysics for modeling stellar structures~\cite{Chandra}) $N(y)=-\gamma y^p$  with $\gamma>0$ and $p\in \mathbb{R}$ an index. Here, we use specifically $p=2$ which leads to a nontrivial but simple solution that was also obtained by G.~Mie in his nonlinear electrodynamics involving solitons~\cite{Mie1912} (see also~\cite{Rosen1965,Schwinger}). The choice $p=2$ has many remarkable properties that can be exploited in the context of the DSP. Writing 
\begin{subequations}
\label{Lane0}
\begin{eqnarray}
U(f^2)=-\frac{r_0^2}{(\frac{g}{4\pi})^4}f^6\label{Lane0a}\\
N(f^2)=-\frac{3r_0^2}{(\frac{g}{4\pi})^4}f^4\label{Lane0b}
\end{eqnarray} 
\end{subequations}
we have in the near field (i.e., Eq.~\ref{ODEnf} with $\mathcal{M}_\Psi(z(\tau))r_0(\tau)\ll 1$) 
\begin{eqnarray}
\boldsymbol{\nabla}^2f(\textbf{x})=-\frac{3r_0^2}{(\frac{g}{4\pi})^4}f^5(\textbf{x}).
\label{Lane}\end{eqnarray} which admits the radial (non topological) soliton
\begin{eqnarray}
f(\textbf{x}):=F(r)=\frac{g}{4\pi}\frac{1}{\sqrt{r^2+r_0^2}}\label{Lanesolution}
\label{Lane2}\end{eqnarray}  (with $r=|\textbf{x}|$) as it can be checked by direct substitution. Eq.~\ref{Lane2} has the asymptotic monopolar limit $F(r)\simeq \frac{g}{4\pi}\frac{1}{r}$ if $r\gg r_0$ and $F(0)=\frac{g}{4\pi r_0}$. $g$ is thus interpreted as a soliton charge (satisfying the integral condition $-\int d^3\textbf{x}N(f^2)f=g$ as shown in Appendix A) while $r_0$ acts as a typical radius for the soliton structure. The static energy of the soliton is  (see Appendix A) $E_s=\frac{g^2}{32r_0}$.  \\ 
\indent The non-linearity Eq.~\ref{Lane0b} is particularly interesting in the context of the DSP since with Eq.~\ref{Lane2} it actually vanishes asymptotically for $r\gg r_0$, i.e., $N(f^2)f=-\frac{3g}{4\pi}\frac{r_0^2}{(r^2+r_0^2)^{\frac{5}{2}}}\rightarrow-\frac{3g}{4\pi}\frac{r_0^2}{r^5}$.
Far-away of the soliton core the monopolar approximation is thus very good and in this limit it is justified to use instead Poisson's equation $\boldsymbol{\nabla}^2f(\textbf{x})=-g\delta^3(\textbf{x})$ for a point-like source. More generally, it is visible that with Eq.~\ref{Lane0} the NLKG equation reduces to  $\Box u(x)\simeq 0$ if the amplitude of $u\rightarrow 0$. It is important to observe that the  asymptotic field $u\propto 1/r$ decays too slowly for applying Ehrenfest theorem (we proved this results in \cite{Submitted}). Therefore we evade the conclusions discussed before for a strongly localized droplet. \\
\indent Eq.~\ref{Lane} possesses an interesting dilation invariance~\footnote{This invariance allows us to circumvent the conclusions of the Hobart-Derrick theorem~\cite{Hobart,Derrick,Goldstone} which usually precludes the existence of static and stable solitons in 3D space. In Appendix~\ref{appDerrick} we give an elementary proof of this result. }. Indeed, it can directly checked   that if $f(\mathbf{x})$ is a solution of Eq.~\ref{Lane} so is the function $\widetilde{f}(\mathbf{x})=\sqrt{\alpha}f(\alpha\mathbf{x})$   where $\alpha\in \mathbb{R}$.  In other words, from Eq.~\ref{Lanesolution}:
\begin{eqnarray}
\widetilde{f}(\mathbf{x})=\widetilde{F}(r)=\frac{\sqrt{\alpha}g}{4\pi}\frac{1}{\sqrt{\alpha^2r^2+r_0^2}}=\frac{g}{4\pi \sqrt{\alpha}}\frac{1}{\sqrt{r^2+\frac{r_0^2}{\alpha^2}}}\label{Lanesolutionscaled}
\end{eqnarray} 
The second expression shows that the new soliton  corresponds to a particle with a new characteristic radius $\tilde{r_0}=\frac{r_0}{\alpha}$ and a new charge $\tilde{g}=\frac{g}{\sqrt{\alpha}}$ ($-\int d^3\textbf{x}N(\widetilde{f}^2)\widetilde{f}=\tilde{g}$), and Eq.~\ref{Lane} can alternatively be written as 
\begin{eqnarray}
\frac{d^2}{dr^2}\widetilde{F}(r)+\frac{2}{r}\frac{d}{dr}\widetilde{F}(r)+\frac{3\widetilde{r_0}^2}{(\frac{\widetilde{g}}{4\pi})^4}\widetilde{F}^5(r)=0
\label{Lanere}\end{eqnarray} with $r_0^2/g^4=\tilde{r_0}^2/\tilde{g}^4$. Furthermore, observe that the quasi-static energy is invariant, i.e. $\widetilde{E}_s=\frac{\tilde{g}^2}{32\tilde{r_0}}=E_s$ during this transformation. Moreover, if $r\gg \tilde{r_0}$ we have the asymptotic field 
  \begin{eqnarray}
\widetilde{F}(r)\simeq \frac{g}{4\pi \sqrt{\alpha}}\frac{1}{r}=\frac{\tilde{g}}{4\pi }\frac{1}{r}\label{LanesolutionscaledFF}
\end{eqnarray} which again confirms  that we have a monopolar quasi-static term corresponding to a charge $\tilde{g}$.\\
\indent Physically, the parameter $\alpha$ can be interpreted as a new collective coordinate for the soliton.  More precisely,  we now assume that during its motion the soliton typical extension $r_0(\tau)$ changes with time $\tau$. We thus write 
\begin{eqnarray}
r_0(\tau):=\tilde{r_0}=r_0(0)/\alpha(\tau),&
g(\tau):=\tilde{g}=\frac{g(0)}{\sqrt{\alpha(\tau)}}
\end{eqnarray} where  $\alpha(\tau)$ defines the dynamics concerning the radius. Therefore, for points located not too far from the soliton center  Eq.\ref{LanesolutionscaledFF} generally describes the field along the hyperplane $\Sigma(\tau)$ of Fig.~\ref{image1}.   In particular, in the near-field  the description is supposed to be very robust because of the condition $\mathcal{M}_\Psi(z(\tau))r_0(\tau)\ll 1$. We thus write in the near-field defined in the proper rest frame $\mathcal{R}_\tau$:
\begin{eqnarray}
F_\tau(r):=\widetilde{F}(r)=\frac{\sqrt{\alpha(\tau)}g(0)}{4\pi}\frac{1}{\sqrt{\alpha(\tau)^2r^2+r_0(0)^2}}\nonumber\\
\label{LanesolutionREscaled}
\end{eqnarray} 
where the parameter $\alpha(\tau)$ now describes the compressibility or `deformability' of the moving soliton droplet defined along the hyperplane $\Sigma(\tau)$.\\ 
\indent The picture obtained is thus the one of a deformable or compressing moving soliton.   It is important to see that $B(\tau)$ and $\alpha(\tau)$ can easily be connected. More precisely, writing the local conservation law $\frac{d}{d\tau}\log{[f^2(z(\tau))\mathcal{M}_\Psi(\tau)\delta^3\sigma_0(\tau)]}=0$ for a fluid element located at the soliton center we have   by integration 
\begin{eqnarray}
f^2(z(\tau))\mathcal{M}_\Psi(\tau)\delta^3\sigma_0(\tau)=f^2(z(0))\mathcal{M}_\Psi(0)\delta^3\sigma_0(0)\nonumber\\ \label{intee}
\end{eqnarray}
Furthermore, from Eq.~\ref{LanesolutionREscaled} $f(z(\tau))=F_\tau(0)=\sqrt{\alpha(\tau)}F_0(0)=\sqrt{\alpha(\tau)}f(z(0)$ and $\delta^3\sigma_0(\tau)=\frac{1}{\alpha^3(\tau)}\delta^3\sigma_0(\tau)$ (a more rigorous justification is given in Appendix C).  Therefore,    Eq.~\ref{intee} leads to 
\begin{eqnarray}
\alpha(\tau)=\sqrt{\frac{\mathcal{M}_\Psi(\tau)}{\mathcal{M}_\Psi(0)}}. \label{inteF}
\end{eqnarray}
Moreover, we also have $\partial v_u(z(\tau))=\frac{d}{d\tau}\log{(\delta^3\sigma_0(\tau))}=\frac{d}{d\tau}\log{(\alpha^{-3}(\tau))}=-\frac{3}{2}\frac{d}{d\tau}\log{(\mathcal{M}_\Psi(\tau))}$ yielding:
\begin{eqnarray}
B(\tau)=\frac{1}{2}\frac{d}{d\tau}\mathcal{M}_\Psi(\tau).\label{inteG}
\end{eqnarray}
Together Eqs.~\ref{inteF} and \ref{inteG} define the complete deformation/compression of the soliton near-field.   In particular,   in the non-relativistic regime where the mass  $\mathcal{M}_\Psi(\tau)$ is approximately constant we have $\alpha(\tau)\simeq 1$ et   $B(\tau)\simeq 0$, i.e., $\partial v_u(z(\tau))\simeq 0$. We thus recover the picture of an incompressible soliton. In the general relativistic case  we get by integration of Eq.~\ref{newd} and the value of $B(\tau)$ the relation 
\begin{eqnarray}
f(z(\tau))=\left(\frac{\mathcal{M}_\Psi(\tau)}{\mathcal{M}_\Psi(0)}\right)^{1/4}f(z(0))\label{inteE}
\end{eqnarray} in agreement with $f(z(\tau))=\sqrt{\alpha(\tau)}f(z(0)$ and Eq.~\ref{inteF}.\\
\indent In the end we thus succeded in obtaining a description of a moving soliton with near-field
\begin{eqnarray}
u(x)=F_\tau(r)e^{i\varphi(x)}\label{summary}
\end{eqnarray}  with $x\in \Sigma(\tau)$ and where $F_\tau$ is given by Eq.~\ref{LanesolutionREscaled}, $\varphi$ by Eq.~\ref{phaseharmony} (phase harmony condition), and the constraints for $\alpha(\tau)$ and $B(\tau)$ are given by Eqs.~\ref{inteF}, \ref{inteG}. The dynamics of the soliton core $z(\tau)$ is piloted  by the guidance formula Eq.~\ref{guidance} and recovers the PWI (e.g., Eq.~\ref{Newton}).
%%%%%%%%%%%
 \section{The Time-symmetric de Broglie double solution  in the far-field}
\label{sec3}  
\indent The previous theory developed for the near-field can be used to define the mide-field and far-field of the soliton. We go back to Eq.~\ref{2c} written as Eq.~\ref{ODE} and dont neglect the mass term $\mathcal{M}_\psi(z(\tau)$. We consider first the  case of an uniform motion  where $\mathcal{M}_\Psi=Const.=\omega_0$ and search for a spherical solution of 
\begin{eqnarray}
\frac{d^2}{dr^2}F(r)+\frac{2}{r}\frac{d}{dr}F(r)+\frac{3r_0^2}{(\frac{g}{4\pi})^4}F^5(r)+\mathcal{M}_\Psi F(r)=0.
\label{Laneradial}\end{eqnarray} In the near-field (i.e., $\mathcal{M}_\Psi r\ll 1$) we have Eq.~\ref{Lanesolution} with asymptotic monopolar limit $F(r)\simeq \frac{g}{4\pi}\frac{1}{r}$.   In the far-field (i.e., $\mathcal{M}_\Psi r\gg 1$) we have $\frac{d^2}{dr^2}F(r)+\mathcal{M}_\Psi F(r)\simeq 0$ which admits the monopole solution:
$F(r)\sim \frac{g}{4\pi}\frac{\cos{(\mathcal{M}_\Psi(z(\tau)) r)}}{r}$. We can easily interpolate these two solutions by writing a solution of Eq.~\ref{Laneradial} as 
$F(r)=\frac{g}{4\pi}\frac{G(r/r_0)}{r}\cos{(\mathcal{M}_\Psi r)}$ we obtain for $G(x)$ the differential equation 
\begin{eqnarray}
\ddot{G}(x)+\frac{3G^5(x)}{x^4}\cos{(\mathcal{M}_\Psi r_0)}-2\mathcal{M}_\Psi r_0\tan{(\mathcal{M}_\Psi r_0)}\dot{G}(x)=0.\label{model}
\end{eqnarray}
After assuming $\mathcal{M}_\Psi(z(\tau))r_0\ll 1$ (i.e., a very small soliton) it reduces to $\ddot{G}(x)+\frac{3G^5(x)}{x^4}=0$ which admits the solution $G(x)=\frac{x}{\sqrt{1+x^2}}$, i.e.,
\begin{eqnarray}
F(r)\simeq\frac{g}{4\pi}\frac{\cos{(\mathcal{M}_\Psi r)}}{\sqrt{r^2+r_0^2}}\label{interpolate1}
\end{eqnarray} which is indeed an interpolation 
 %%%%%%%%%%%%%%%%
\begin{figure}[h]
\centering
\includegraphics[width=7 cm]{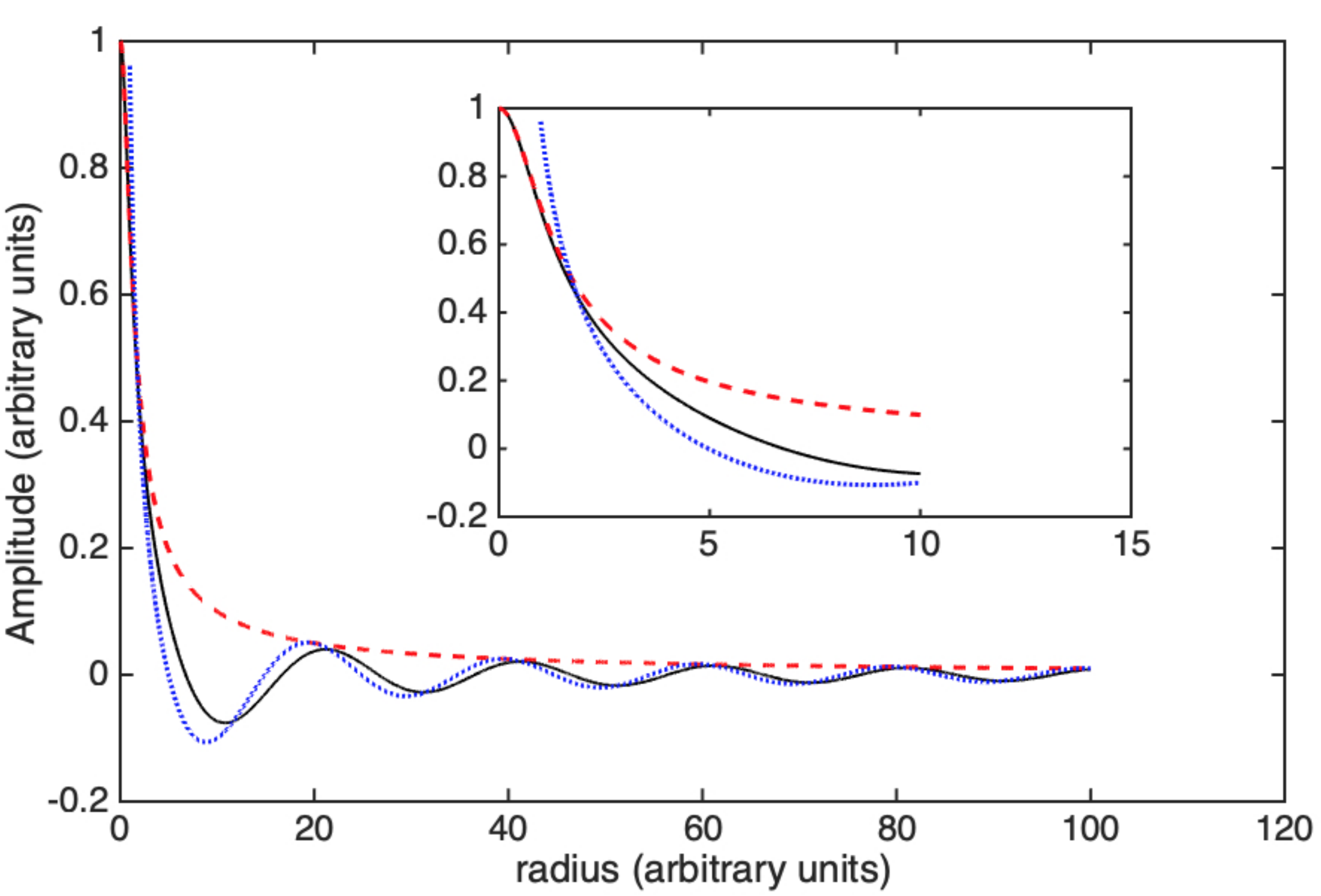}
\caption{Soliton profile $F(r)$ as a function of the radius $r$. The black curve shows the solution $F(r)$ of the equation $\Delta F=-3F^5-A F$ with  $A=0.1$. The red dashed curve corresponds to the case $A=0$ (i.e., the Lane-Emden soliton $F(r)=\frac{1}{\sqrt{r^2+1}}$). The solutions are obtained numerically by imposing $F(0)=1$ and $\frac{d}{dr}F(0)=0$.   The blue dooted curve corresponds to the asymptotic stationary monopole field $F(r)=\frac{\cos{(\sqrt{A} r)}}{r}$. The inset in the right upper corner of the figure is a zoom of the three curves near the soliton center.
}    \label{image2}
\end{figure}
%%%%%%%%%%%%%%%%%%%%%%%%%%%%%%%%%%%%%%   
between the monopolar and the Lane-Emden quasi-static solutions (see Fig.~\ref{image2} for a numerical calculation). We check that $-\int d^3\textbf{x}N(F^2)F=g+gO((\mathcal{M}_\Psi r_0)^2)\simeq g$.   \\
\indent We stress that already in 1925~\cite{deBroglie1925a,deBroglie1925b} de Broglie using  the linear  d'Alembert equation $\Box u(x)=0$ developed  a  preliminary version of  the DSP  admitting the monopolar singular field 
\begin{eqnarray}
u(t,r)=\frac{g}{4\pi}e^{-i\omega_0 t}\frac{\cos{(\omega_0 r)}}{r}.\label{debroglie1}
\end{eqnarray} This $u-$field  for a free particle in uniform motion is seen in the Lorentz-reference frame where the particle is at rest at the origin.  Eq.~\ref{debroglie1} is actually a singular solution of the inhomogeneous d'Alembert equation $\Box u(t,\textbf{x})=g\delta^{3}(\textbf{x})e^{-i\omega_0 t}$ and merges with the far-field of our soliton Eq.~\ref{interpolate1} and $u=F(r)e^{-i\omega_0 t}$ if $\omega_0:=\mathcal{M}_\Psi$.\\ 
\indent We stress that Eq.~\ref{debroglie1} reads $gG^{(0)}_{sym,\omega_0}(r)e^{-i\omega_0 t}$ where
\begin{eqnarray}G^{(0)}_{sym,\omega}(R)=\frac{\cos{(\omega R)}}{4\pi R}=\frac{1}{2}[\frac{e^{i\omega R}}{4\pi R}+\frac{e^{-i\omega R}}{4\pi R}]\label{Greenfreq}\end{eqnarray}  ($R=|\mathbf{x}-\mathbf{x}'|$) is the time-symmetric Green function\footnote{We have $G^{(0)}_{sym,\omega}(R)=\frac{1}{2}[G^{(0)}_{ret,\omega}(R)+G^{(0)}_{adv,\omega}(R)]$ and $G^{(0)}_{ret/adv,\omega}(R)=\frac{e^{\pm i\omega R}}{4\pi R}$ are the retarded and advanced Green functions respectively.} of the Helmholtz equation: $[\omega^2 +\boldsymbol{\nabla}^2]G^{(0)}_{\omega}(R)=-\delta^3(\mathbf{x}-\mathbf{x}')$. In other words, $u(t,r)$ in Eq.~\ref{debroglie1} is a time-symmetric solution $u=\frac{1}{2}(u_{ret}+u_{adv})$ of  $\Box u(t,\textbf{x})=g\delta^{3}(\textbf{x})e^{-i\omega_0 t}$. This is fundamental because it leads to the stability of the micro-object: The energy radiation losses  associated with the retarded wave are exactly compensated by the energy flow associated with  the converging  advanced wave. Furthermore, it implies a time-symmetric causality which is reminiscent of early ideas by Tetrode and Page~\cite{Tetrode,Page} for explaining the stability of atomic orbits. Such ideas were later resurrected by Fokker~\cite{Fokker}, Feynman and Wheeler in their absorber theory~\cite{WF}, and by Hoyle and Narlikar for cosmological models involving a time-symmetric creation-field \cite{Hoyle}. Interestingly, this idea involving time-symmetry was also discussed in 1925 by de Broglie~\cite{deBroglie1925a,deBroglie1925b} but was soon abandoned by him and he never came back to this suggestion (even after his collaborator Costa de Beauregard developped a retrocausal interpretation of the EPR paradox~\cite{Beauregard}). \\
%%%%%%%%%%%%%%%%
\begin{figure*}[h]
\centering
\includegraphics[width=12 cm]{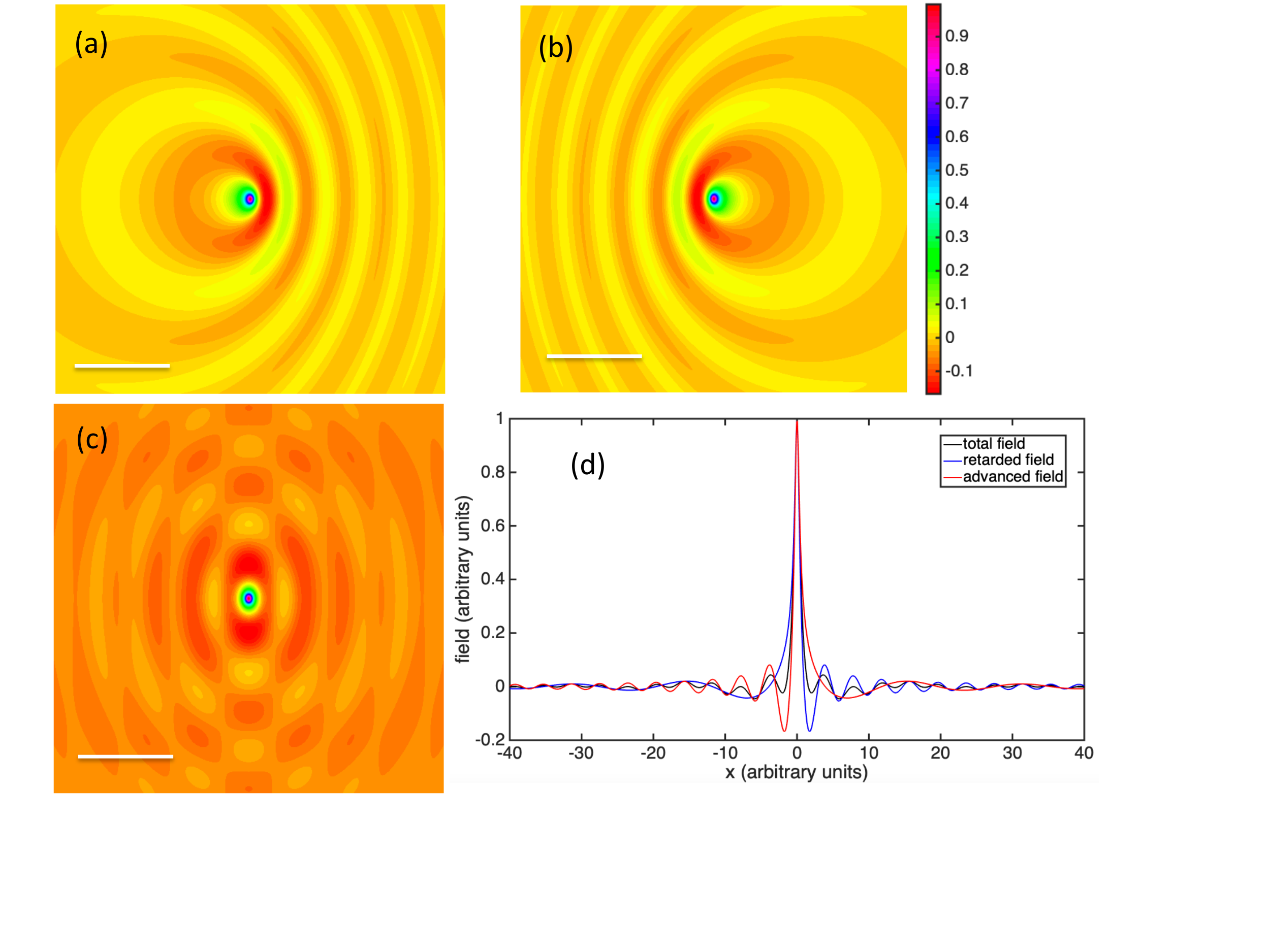}
\caption{Soliton profile $\textrm{Re}[u(x,y,z=0,t=0)]$ for a particle in  uniform motion (velocity $v_x=0.6$ along the $x$ direction) in the laboratory frame. (a) and (b) are respectively  the advanced and retarded fields contributions   $\textrm{Re}[u_{ret./adv.}(x,y,z=0,t=0)]$, and (c) is the whole time-symmetric field $\textrm{Re}[u_{sym}(x,y,z=0,t=0)]=\frac{1}{2}\textrm{Re}[u_{ret.}(x,y,z=0,t=0)+u_{adv.}(x,y,z=0,t=0)]$. (d) shows a crosscut along the x direction of the three fields $\textrm{Re}[u_{ret.}(x,y=0,z=0,t=0)]$ (red curve), $\textrm{Re}[u_{adv.}(x,y=0,z=0,t=0)]$ (blue curve), and $\textrm{Re}[u_{sym.}(x,y=0,z=0,t=0)]$ (black curve). For the present calculations we used   $\omega_0\gamma=1$ and $a=\frac{1}{\sqrt{10}}\gamma\simeq0.3953$ in Eqs.~\ref{sym}, \ref{retadv}. The scale bar used in (a-c)  is 10 units  long.
}    \label{image3}
\end{figure*}
%%%%%%%%%%%%%%%%%%%%%%%%%%%%%%%%%%%%%%    
\indent To appreciate the time-symmetric nature of our soliton  we represent in Fig.~\ref{image3}  $\textrm{Re}[u(x,y,z=0,t=0)]$ in the laboratory frame where the soliton moves at the velocity $v_x$ along the $+x$ direction. Using Eq.~\ref{interpolate1} for the interpolated solution  we have 
\begin{eqnarray}
\textrm{Re}[u(\mathbf{x},t)]\simeq \frac{g}{4\pi}\frac{\cos{(\omega_0 R)}cos{[\omega_0\gamma(t-v_xx)}]}{\sqrt{r_0^2+R^2}}\label{sym}
\end{eqnarray} with $R=\sqrt{(x-v_x t)^2\gamma^2+y^2+z^2}$ and $\gamma=(1-v_x^2)^{-\frac{1}{2}}$. Moreover,  writing $u:=u_{sym}=\frac{1}{2}[u_{adv.}+u_{ret.}]$ and still using the interpolated field we can define 
\begin{eqnarray}
\textrm{Re}[u_{ret./adv.}(\mathbf{x},t)]\simeq \frac{g}{4\pi}\frac{\cos{[\omega_0\gamma(t-v_xx)\mp \omega_0R]}}{\sqrt{r_0^2+R^2}}\label{retadv}
\end{eqnarray} associated with propagating diverging/converging waves.  Comparing Eqs.~\ref{retadv} and \ref{sym} shows that the particle core is surfing a wave front reminiscent of what is occurring  with a airplane in the subsonic regime (with here the velocity of light replacing the velocity of sound).  The  wave front precedes the particle in the retarded case and follows the particle in the advanced case.  The superposition $\textrm{Re}[u_{sym.}(x,y,z=0,t=0)$ induces a phase wave  $cos{[\omega_0\gamma(t-v_xx)}]$ associated with de Broglie's guiding field (i.e., the guiding wave involved in the PWI).\\
%%%%% 
 \begin{figure}[h]
\centering
\includegraphics[width=8cm]{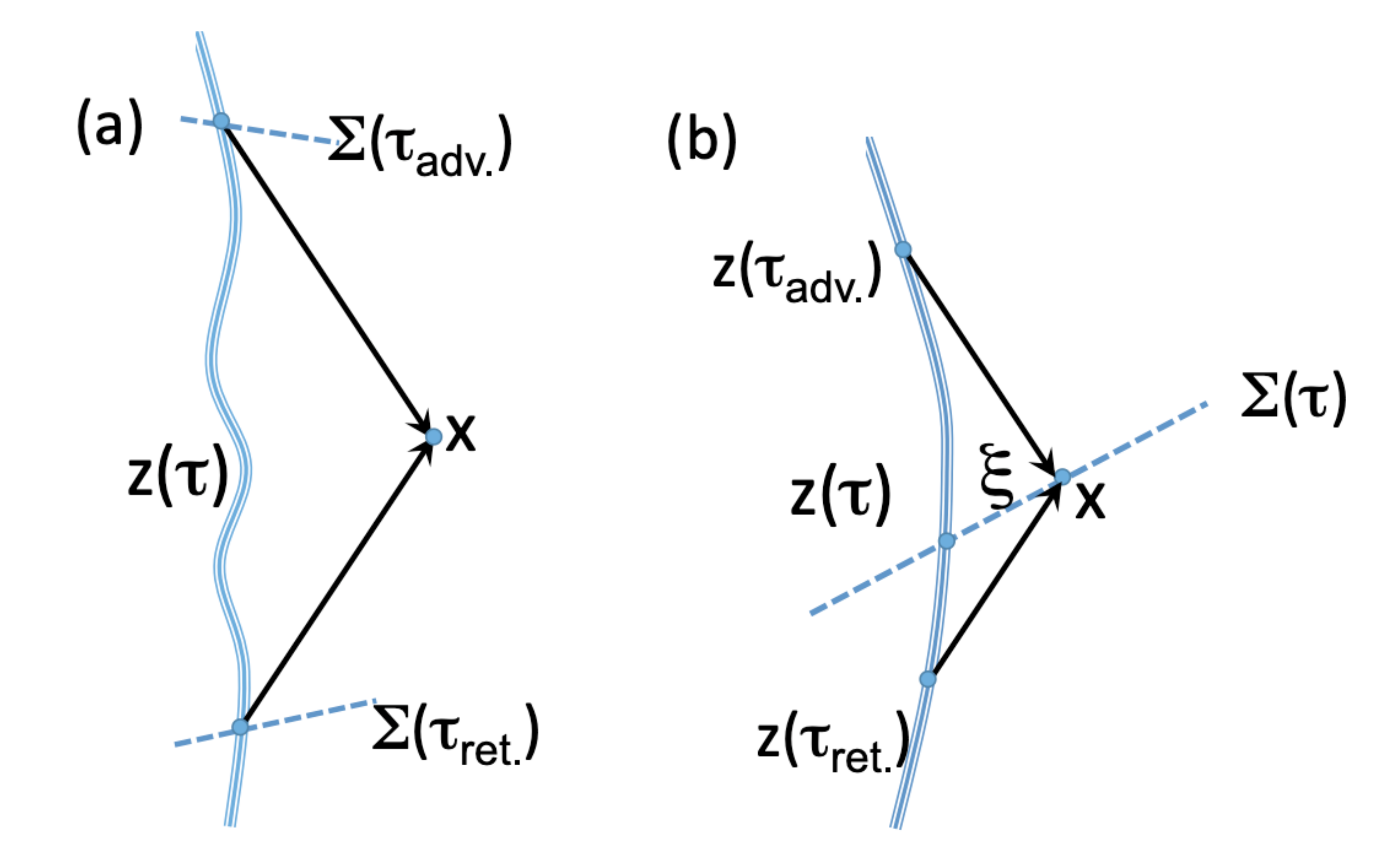}
\caption{(a) Far-field of the soliton $u(x)$ at point $x$ away from the `singularity' trajectory $z(\tau)$. The field is emitted at retarded/advanced (proper) time $\tau_{ret/adv}$. (b) Same as in (a) but in the vicinity of the trajectory $x\sim z$ where we evaluate the field in the hyperplane $\Sigma(\tau)$  at distance $\xi=x-z(\tau)$.
}    \label{image4}
\end{figure} 
%%%%
\indent The previous theory for the far-field can be generalized. For this purpose start from Eq.~\ref{1b}  written as  $D^2u(x)=-N(u^\ast (x)u(x))u(x):=J(x)$. Using  Green's theorem a formal solution reads \begin{eqnarray}
 u(x)=\int K(x,y)J(y)d^4y+ u_{free}(x)  \label{debrogliefarfield}
\end{eqnarray} where $u_{free}(x)$ is a solution of the homogeneous equation $D^2u(x)=0$, and the propagator $K(x,x')$ satisfies $D^2 K_{sym}(x,x')=\delta^{4}(x-x')$. Consider first the case  $A(x)=0$ (i.e., absence of external field). We see that a natural choice corresponds to $u_{free}(x)=0$ and $K^{(0)}_{sym}(x,x')=\frac{K^{(0)}_{ret}(x,x')+K^{(0)}_{adv}(x,x')}{2}$    
where \begin{eqnarray}
K^{(0)}_{sym}(x,x')=\frac{\delta[(x-x')^2)]}{4\pi}=\frac{1}{2}[\frac{\delta(t-t'-R)}{4\pi R}+\frac{\delta(t-t'+R)}{4\pi R}]\label{green0}
\end{eqnarray} (with $R=|\mathbf{x}-\mathbf{x}'|^2$) is the time-symmetric propagator\footnote{We have also $K^{(0)}_{sym}(x,x')=\int_{-\infty}^{+\infty}G^{(0)}_{sym,\omega}(R)e^{-i\omega(t-t')}\frac{d\omega}{2\pi}$ with $G^{(0)}_{sym,\omega}(R)$ given by Eq.~\ref{Greenfreq}}.\\
\indent As shown in Fig.~\ref{image4}(a) the evaluation of the far-field at point $x$ requires the knowledge of the source term $J(x)$ in the vicinity of the trajectory $z(\tau)$ along two hyperplanes  $\Sigma(\tau_{ret.})$, $\Sigma(\tau_{adv.})$ associated with retarded and advanced emissions by the particle. These planes are obtained by finding  intersections of the trajectory $z(\tau)$ with the backward and forward light cones with common apex located at point $x$. We have approximately:
\begin{eqnarray}
u(x)\simeq \frac{1}{2}\int\int_{\Sigma(\tau_{ret.})} K^{(0)}_{ret}(x,z(\tau_{ret}))J(\tau_{ret},\boldsymbol{\xi})d^3\boldsymbol{\xi}d\tau_{ret.}
\nonumber\\+\frac{1}{2}\int\int_{\Sigma(\tau_{adv.})} K^{(0)}_{adv}(x,z(\tau_{adv}))J(\tau_{adv},\boldsymbol{\xi})d^3\boldsymbol{\xi}d\tau_{adv.}\nonumber\\
\end{eqnarray}  
where the integrations are done  in the local rest frame $\mathcal{R}_{\tau_{ret.}}$ and $\mathcal{R}_{\tau_{adv.}}$. After spatial integration over the hyperplanes $\Sigma(\tau_{ret.})$, $\Sigma(\tau_{adv.})$ and using the lowest order approximation  $\int_{\Sigma(\tau)}J(\tau,\boldsymbol{\xi})d^3\boldsymbol{\xi}= -\int d^3\textbf{x}N(F_\tau^2)F_\tau e^{iS(z(\tau))}\simeq g(\tau)e^{iS(z(\tau))}=g(0)\frac{e^{iS(z(\tau))}}{\sqrt{\alpha(\tau)}}$ (see Eq.~\ref{summary} with $\varphi\simeq S$) we obtain for the far-field:
\begin{eqnarray}
 u(x)=g(0)\int_{(C)}K^{(0)}_{sym.}(x,z(\tau))\frac{e^{iS(z(\tau))}}{\sqrt{\alpha(\tau)}}d\tau \label{debroglieSing2AS}
\end{eqnarray} with $\alpha(\tau)=\sqrt{\frac{\mathcal{M}_\Psi(\tau)}{\mathcal{M}_\Psi(0)}}$ and the integration is over the whole trajectory $C$. This field is a solution of 
\begin{eqnarray}
\Box u(x)=g(0)\int_{(C)}\delta^{4}(x-z(\tau))\frac{e^{iS(z(\tau))}}{\sqrt{\alpha(\tau)}}d\tau\nonumber\\=g(0)\delta^{3}(\mathbf{x}-\mathbf{z}(t))\frac{e^{iS(t,\mathbf{z}(t))}}{\sqrt{\alpha(\tau)}}\sqrt{1-\mathbf{v}^2(t)} \label{debroglieSingbis}
\end{eqnarray} which shows that the coupling vanishes when the velocity of the particle $\mathbf{v}$ approaches the celerity of light.\\ 
\indent We stress that using Eq.~\ref{summary} in $\int_{\Sigma(\tau)}J(\tau,\boldsymbol{\xi})d^3\boldsymbol{\xi}$ looks like a physical ansatz. To justify the self-consistency of the ansatz we now explicit Eq.~\ref{debroglieSingbis} using Eq.~\ref{green0}: 
 \begin{eqnarray}
 u(x)=\frac{1}{2}\left([\frac{g(\tau)e^{iS(z(\tau))}}{4\pi\rho(\tau)}]_{\tau_{ret.}}+[\frac{g(\tau)e^{iS(z(\tau))}}{4\pi\rho(\tau)}]_{\tau_{adv.}}\right)\label{Lienard}
 \end{eqnarray} with $\rho(\tau)=|(x-z(\tau))\cdot \dot{z}(\tau)|$. The derivation of this formula is clearly reminiscent from the Lienard-Wiechert potentials in classical electrodynamics \footnote{We mention that F.~Fer in 1957 developed a method for analyzing the motion of singularities in the context of the DSP~\cite{Fer1957,Fer1973}. Moreover, his approach using only retarded Green's functions missed the time-symmetry needed to recover the wave particle duality considered here.}. To justify the ansatz used above the goal is  to compute  $u(x)$ with Eq.~\ref{Lienard} in the vicinity of  the trajectory $z(\tau)$ in order to recover the asymptotic near-field $u\propto 1/r$ obtained in Sec.~\ref{sec2}. The method has been already developed by Dirac \cite{Dirac} for the classical electron and requires lengthy calculations that will not be shown here for questions of space. We here summarize the main steps of the methods. As shown in Fig.~\ref{image4}(b) the field is evaluated in the hyperplane $\Sigma(\tau)$  at a distance $r=\sqrt{-\xi^2}$ that requires the retarded and advanced fields in Eq.~\ref{Lienard} at (proper) times $\tau_{ret}=\tau-\sigma_-$ and  $\tau_{ret}=\tau+\sigma_+$ with\footnote{More precisely we have $\sigma_\mp=r(1+\frac{\xi \ddot{z}}{2} \mp \frac{r\xi z^{(3)}}{2}-\frac{r^2(\ddot{z})^2}{24}+\frac{3(\xi \ddot{z})^2}{8})+O(r^4)$.} $\sigma_{\pm} \sim r$.  Using the conditions $(x-z(\tau\mp \sigma_{\mp}))^2=0$ for the points on a light cone we deduce after lengthy calculations \cite{Dirac} the values $\rho_{\mp}=\rho(\tau\mp \sigma_\mp)$, i.e.:
 \begin{eqnarray}
 \frac{1}{\rho_\mp}=\frac{1}{r}[1+\frac{1}{2}\xi\ddot{z}+\frac{3}{8}(\xi \ddot{z})^2 \mp \frac{r}{3}\xi z^{(3)}+\frac{5r^2(\ddot{z})^2}{24}+O(r^3)]
\end{eqnarray}  where the derivatives  and $\xi$ are calculated at time $\tau$.  The field is thus $u(x)=\frac{u_-+u_+}{2}$ with $u_{\mp}=\frac{g(\tau\mp \sigma_{\mp})e^{iS(\tau\mp \sigma_{\mp})}}{8\pi \rho_{\mp}}$  that leads to:
\begin{eqnarray}
u(x)=\frac{g(\tau)e^{iS(z(\tau))}}{4\pi r}[1+\frac{\xi\ddot{z}}{2}+\frac{r^2}{2}(i\ddot{S}-(\dot{S}-i\frac{\dot{g}}{g})^2)\nonumber\\+r^2\frac{d^2}{d\tau^2}\ln{(g)}+\frac{3}{8}(\xi\ddot{z})^2+\frac{5}{24}(r\ddot{z})^2+O(r^3)].\label{asymp}
\end{eqnarray} for points $x\in\Sigma(\tau)$ at a distance $r$ from $z(\tau)$.\\ 
\indent From Eq.~\ref{asymp} we deduce that at the lowest order we have indeed $u(x)\simeq\frac{g(\tau)e^{iS(z(\tau))}}{4\pi r}$  and we recover the asymptotic soliton near-field discussed in Sec.~\ref{sec2}.  This shows that our ansatz concerning $\int_{\Sigma(\tau)}J(\tau,\boldsymbol{\xi})d^3\boldsymbol{\xi}$ is indeed justified.  Moreover,   in the case of an uniform motion with $\ddot{z}=0$, $\ddot{S}=0$, $\dot{S}=-\omega_0$, $\dot{g}=0$, $\ddot{g}=0$ we have $u(x)=\frac{ge^{-i\omega_0\tau}}{4\pi r}[1-\frac{\omega_0r^2}{2}+O(r^3)]\simeq\frac{ge^{-i\omega_0\tau}\cos{(\omega_0 r)}}{4\pi r}$ that is the field associated with the monopole discussed above.\\
\indent Furthermore, from Eq.~\ref{asymp} and the definition $u=fe^{i\varphi}$ we compute the ratio $\frac{u(x)}{u^\ast(x)}=e^{i2\varphi(x)}$ and obtain
\begin{eqnarray}
\frac{u(x)}{u^\ast(x)}=e^{i2\varphi(x)}=e^{i2S}[1+ir^2(\ddot{S}+2\frac{\dot{S}\dot{g}}{g})+O(r^3)].
\end{eqnarray} Comparing this with the Taylor expansion $e^{i2\varphi(x)}=e^{i2\varphi(z)}[1+i2\xi\partial\varphi(z)+O(r^2)]$ we deduce immediately that the first-order term must vanish 
 \begin{eqnarray}
\xi\cdot\partial\varphi(z)=0
\end{eqnarray}  and since we have also by definition $\xi\dot{z}=0$ we have thus $\dot{z}(\tau)$ parallel (i.e., proportional) to $\partial \varphi (z)$. In other words, since $\dot{z}^2=1$, we recover the guidance formula Eq.~\ref{guidance} $\dot{z}(\tau)=-\frac{\partial\varphi(z(\tau))}{\sqrt{(\partial\varphi(z(\tau)))^2}}$ (in absence of external field $A(x)$)  discussed in Sec.~\ref{sec2} in the near-field. The fact that we can recover this result from Eq.~\ref{Lienard} associated with the far-field again confirms the self-consistency of our approach.\\
\indent In presence of an external field $A(x)\neq 0$ the previous propagator method can be generalized. For this we first replace the partial derivative $\partial$ by the covariant derivative $D=\partial +ieA(x)$ in Eq.~\ref{debroglieSingbis} leading to   
\begin{eqnarray}
D^2 u(x)=g(0)\int_{(C)}\delta^{4}(x-z(\tau))\frac{e^{iS(z(\tau))}}{\sqrt{\alpha(\tau)}}d\tau\label{debroglieSingtri}
\end{eqnarray} with solution
\begin{eqnarray}
 u(x)=g(0)\int_{(C)}K_{sym}(x,z(\tau))\frac{e^{iS(z(\tau))}}{\sqrt{\alpha(\tau)}}d\tau \label{debroglieSing3AS}
\end{eqnarray} and $K_{sym}(x,x')$ the time-symmetric propagator solution of $D^2 K_{sym}(x,x')=\delta^{4}(x-x')$. A formal solution for the propagator reads \begin{eqnarray}
K_{sym}(x,x')=K^{(0)}_{sym}(x,x')+K^{(ref)}_{sym}(x,x')\label{Greentotal}
\end{eqnarray} where $K^{(ref)}_{sym}(x,x')=\int d^4yK^{(0)}_{sym.}(x,y)\hat{\mathcal{O}}_y K_{sym}(y,x')$ (with the operator $\hat{\mathcal{O}}_y:=e^2A(y)^2-ie\partial_y A(y)-2ieA(y)\partial_y$) defines the reflected part of the propagator resulting from the interaction of the vacuum solution $K^{(0)}_{sym}$ with the potential $A$. Therefore, the $u-$field splits as \begin{eqnarray}
u(x)=u^{(0)}(x)+u^{(ref)}(x)
\end{eqnarray} where  $u^{(0)}(x)$ evaluated using $K^{(0)}_{sym}(x,x')$ leads to Eq.~\ref{Lienard} and $u^{(ref)}(x)$ is evaluated using $K^{(ref)}_{sym}(x,x')$. As we saw the singular field $u^{(0)}(x)$ is diverging as $u(x)\simeq\frac{g(\tau)e^{iS(z(\tau))}}{4\pi r}$ near $x\sim z(\tau)$. The reflected part is in general a much more regular and weaker field near $x\sim z(\tau)$.  \\
\indent Moreover, near the point $x\sim z(\tau)$ we can assume $A(x)\simeq A(z)=const.$ (the soliton is supposed much smaller than the variation of $A$) and we check directly\footnote{A proof is obtained by using the Fourier transform $K(x,x')=\int\frac{d^4k}{(2\pi)^4}e^{ik(x-x')}G_k$. Eq.~\ref{Greentotal} reads thus $G_k=G_k^{(0)}+G_k^{(0)}(e^2A^2-2eAk)G_k$, i.e., $G_k=\frac{G_k^{(0)}}{1-(e^2A^2-2eAk)G_k^{(0)}}$. Using $G_k^{(0)}=-1/k^2$ we deduce  $G_k=-1/(k-eA)^2$ and after using the inverse Fourier transform we obtain $K_{sym}(x,x')=K^{(0)}_{sym}(x,x')e^{-ieA(z)(x-x')}$.} that the function $K_{sym}(x,x')=K^{(0)}_{sym}(x,x')e^{-ieA(z)(x-x')}$ is a solution of  $(\partial+ieA(z))^2 K_{sym}(x,x')=\delta^{4}(x-x')$. Inserting this result in Eq.~\ref{debroglieSing2AS} where $K_{sym}(x,z(\tau))$ replaces $K^{(0)}_{sym}(x,z(\tau))$ we obtain once more Eq.~\ref{Lienard} with the substitution $e^{iS(z(\tau_{ret/adv}))}\rightarrow e^{iS(z(\tau_{ret/adv}))}e^{-ie(x-z(\tau_{ret/adv})A}$.  In the vicinity of $z(\tau)$ in the hyperplane $\Sigma(\tau)$ we obtain at the lowest order:
\begin{eqnarray}
\frac{u(x)}{u^\ast(x)}=e^{i2S}[1-i2e\xi A(z(\tau))+O(r^2)]=e^{i2\varphi(z)}[1+i2\xi\partial\varphi(z)+O(r^2)],
\end{eqnarray} which implies 
\begin{eqnarray}
\xi\cdot (\partial\varphi(z)+eA(z))=0.
\end{eqnarray} Therefore, using once more $\xi\dot{z}=0$,  we recover the guidance formula Eq.~\ref{guidance}, i.e., $\dot{z}(\tau)=-\frac{\partial\varphi(z(\tau))+eA(z(\tau))}{\sqrt{(\partial\varphi(z(\tau))+eA(z(\tau)))^2}}$ derived in Sec.~\ref{sec2}.\\ 
\indent This analysis shows that even if in general  $|u^{(0)}(x)|\gg |u^{(ref)}(x)|$ in the vicinity of $x\sim z$
 the phase is however  globally influenced by the presence of the external field $A(x)$ imposing the guidance formula.
 %%%%%%%%%%%
  \section{Discussion: Entanglement, generalizations, perspectives}
\label{sec4}  
\indent In order to conclude this article we would like to emphasize some general properties of our model. First, concerning the methodology we started in Sec.~\ref{sec2} with a near-field approach assuming a field with a spherical symmetry $f=F_\tau(r)$ (see Eq.~\ref{LanesolutionREscaled}). This actually neglects the contribution of the reflected field. The consistency of our model becomes more obvious if we formally write the full $u-$field as $u(x)=u^{(0)}(x)+u^{(ref)}(x)$ with 
\begin{eqnarray}
u^{(0/ref)}(x)=-\frac{3r_0^2}{(\frac{g}{4\pi})^4}\int K^{(0/ref)}_{sym}(x,y)(f(y))^5e^{i\varphi(y)}d^4y\nonumber\\
\simeq-\frac{3r_0^2}{(\frac{g}{4\pi})^4}\int K^{(0/ref)}_{sym}(x,y)(f^{(0)}(y))^5e^{i\varphi(y)}d^4y
\end{eqnarray}  where we used the approximation $f\simeq f^{(0)}$ for evaluating the source term in the second line. This is justified since  we assume $f^{(ref)}\ll f^{(0)}$ in the core region of the soliton where the integral contributes.  This shows that  $u^{(0/ref)}(x)$ are determined  by the knowledge of the soliton $f^{(0)}(x)$ in the near-field as assumed in Section. \ref{sec2}.   Moreover we could in principle obtain deviations to this approximation. That could occur for regimes where the reflected field is not small, e.g., in very strong (relativistic) fields leading to further non-linearities.\\ 
\indent To give an illustration of this issue consider a particle at rest in the middle of a spherical ideal cavity of  radius $R$ with a perfectly reflecting wall associated with an infinite potential wall. The far-field stationary spherical solution of the equation  $\Box u(t,\textbf{x})=g\delta^{3}(\textbf{x})e^{-i\omega t}$  reads 
\begin{eqnarray}
u(t,r)=\frac{ge^{-i\omega t}}{4\pi r}[\cos{(\omega r)}-\textrm{cotan}(\omega R)\sin{(\omega r)}]\label{cavity}
\end{eqnarray} and obeys the boundary condition $u(R,t)=0$. Near the origin (where $\omega r<<1$) the reflected field  $f^{(ref)}=\textrm{cotan}(\omega R)\frac{g\sin{(\omega r)}}{4\pi r}\simeq -\textrm{cotan}(\omega R)\frac{g\omega }{4\pi}$  is in general much smaller than  $f^{(0)}=\frac{g\cos{(\omega r)}}{4\pi r}\simeq \frac{g}{4\pi r} $  unless    the `cotan' term is diverging  which occurs if $\omega=\frac{m\pi}{R}$ with $m\in \mathbb{N}$. If that happens then the field in the cavity blows up and the approximations $f^{(ref)}\ll f^{(0)}$ breaks down.   Moreover,  don't forget that the particle is actually guided by the LKG equation   Eq.~\ref{1c} with spherical eigen-solutions $\Psi_n(t,r)=\frac{\sin{(n\pi r/R)}}{r} e^{-i\omega_n t}$ with $\omega_n=\sqrt{(\omega_0^2+\frac{n^2 \pi^2}{R^2})}\simeq \omega_0+\frac{n^2 \pi^2}{2\omega_0R^2}$ and $n\in \mathbb{N}$. The particle is at rest in agreement with the PWI and we have $\omega=\omega_n\simeq \omega_0$ in Eq.~\ref{cavity}. We see that problem occurs only if $\omega_0\simeq \frac{m\pi}{R}$. But this possibility can be rejected for at least two reasons. First, this would imply a strong conspiracy or fine-tuning where the Compton wavelength of the particle $\lambda_0=2\pi/\omega_0$ matches $2R/m$. This corresponds to  very small cavities of the size of the particle  and we enter in the QED regime where particle/antiparticle pairs could be created.  This regime is not considered in our analysis.   
%%%%%%%%%%%%%%%%
\begin{figure}[h]
\centering
\includegraphics[width=12 cm]{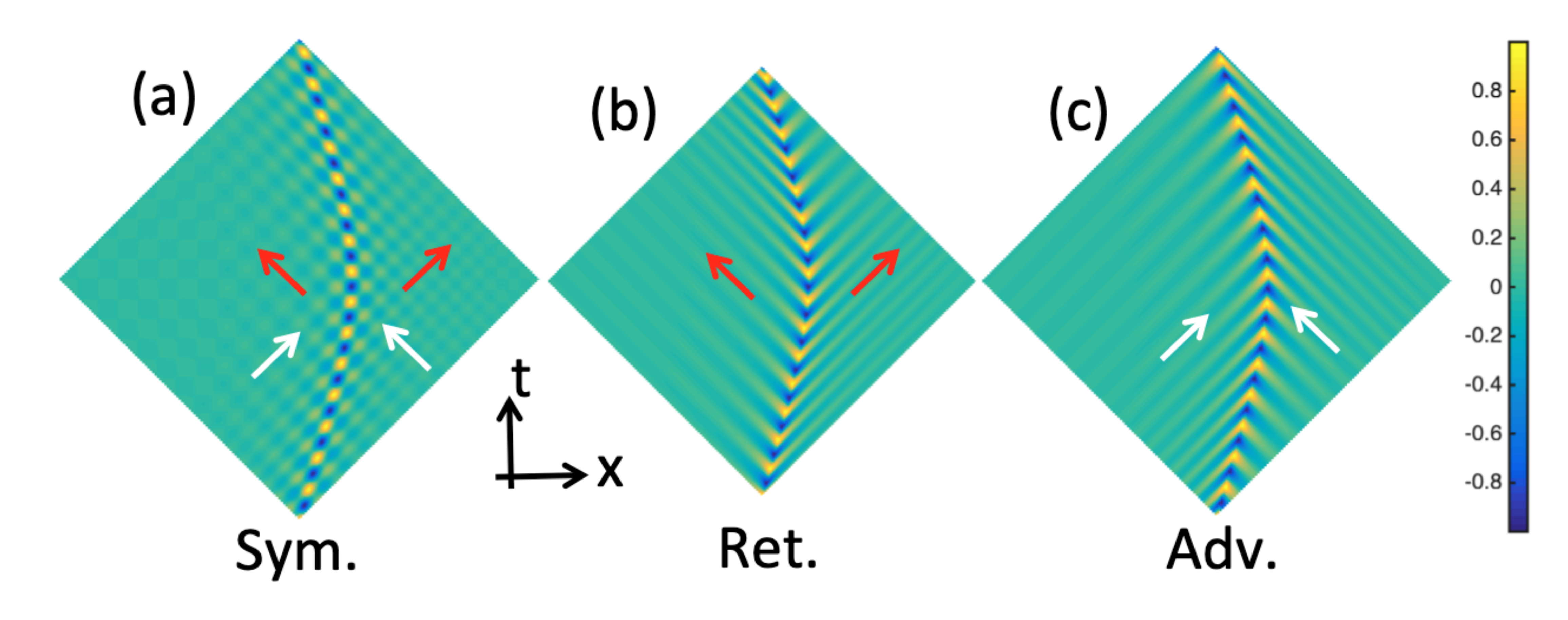}
\caption{(a) Time-symmetric $u-$field associated with a particle following a curved trajectory in the plane $x-t$ [here we use the hyperbolic trajectory $x(t)=-\sqrt{x_0^2+v_0^2t^2}$]. (b) and (c) show the retarded and advanced $u-$fields associated wit the same trajectory. The fields are calculated numerically using expressions Eqs.~\ref{debroglieSing2AS},\ref{Lienard} and similar ones for retarded/advanced fields. The singularity of these far-fields on the particle are removed by modifying the pole as $\frac{1}{r}\rightarrow \frac{1}{\sqrt{(r^2+r_0^2)}}$ with  $r_0$ a small constant. For simplicity we here used the classical phase $S=-\omega_0\tau$ for the particle action with $\tau$ the proper time.}    \label{image5}
\end{figure}
%%%%%%%%%
 Moreover, the second more physical reason for rejecting this implausible resonance is that in general the potential barrier is not infinite and we can show that if the potential  step $|eV|$ is smaller than $\omega_0$  the field given by Eq.~\ref{cavity} is modified: There is no strong reflectivity at the boundary   $r=R$ and $u^{(0)}$ is mostly unaffected~\footnote{To prove this  rather general statement a qualitative argument could go like this: Considering  the LKG equation  $D^2\Psi=-\omega_0^2\Psi$ in a electrostatic potential $V(\mathbf{x})$  the first Born order scattering amplitude  for an incident plane wave  $\Psi_0(\mathbf{x})=e^{ik\mathbf{n}_0\cdot \mathbf{x}}$ (with $k^2=\omega^2-\omega_0^2$) reads $\Psi_s\simeq-2\omega\int d^3\mathbf{x}''G_\omega^{(0)}(\mathbf{x},\mathbf{x}'')eV(\mathbf{x}'')\Psi_0(\mathbf{x}'')\simeq -2\omega\frac{e^{ikr}}{4\pi r} e\hat{V}_{\mathbf{q}}$ where we neglected the quadratic term  $e^2V^2$, $G_\omega^{(0)}(\mathbf{x},\mathbf{x}'')=\frac{e^{ikR|\mathbf{x}-\mathbf{x}''|}}{4\pi|\mathbf{x}-\mathbf{x}''|}$ [computed here for a retarded wave], and where $\hat{V}_{\mathbf{q}}$ is the Fourier transform of the potential at the wave wavevector $\mathbf{q}=k(\mathbf{n}_s-\mathbf{n}_0)$.  In a Coulomb field for example we have $\Psi_s\simeq -\frac{2\omega \alpha}{\mathbf{q}^2}\frac{e^{ikr}}{ r}$. The same calculation done for a plane wave solution of the linearized equation for $u$ $D^2u=0$ leads to the same expression with $\omega^2$ replacing $k^2$. Therefore the scattered field  $u_s$ is smaller than $\Psi_s$ by a coefficient $\frac{u_s}{\psi_s}\simeq k^2/\omega^2=v^2$ where $v$ is the particle velocity.   In general $v^2\ll 1$ and $u_s$ is negligible.}, i.e., $u\simeq u^{(0)}$.  Again all this analysis is consistent if the potential is not too strong so that particle/antiparticle pairs are not generated (pairs are potentially generated if $|eV| \simeq 2 \omega_0$).\\
\indent  An important related  problem concerns energy conservation and causality for a particle moving in an external field.  Consider a particle following a curved trajectory like the one shown in Fig.~\ref{image5} and emitting a retarded + advanced field as given in Eqs.~\ref{debroglieSing2AS},\ref{Lienard}, i.e., neglecting the reflected part  for simplicity. As visible on Fig.~\ref{image5}(a) the time-symmetric field implies that waves are constantly radiated into the future and into the past directions.  De Broglie in  1957 \cite{deBroglie1956} analyzed the problem in terms of retarded waves (as shown in Fig.~\ref{image5}(b)) and concluded that the basic DSP leads to a paradox known as Perrin's objection (for a discussion see \cite{Drezet1}): Following this objection a particle interacting  with an external  field, like a beam splitter, should radiate  energy in empty branches not followed by the particle. After several interactions of that kind the particle (i.e., the $u-$wave) should have lost all its energy in contradiction with experiments showing that particles are detected with a finite energy (the same issue remains in the double-slit experiment where the potential acting on the particle is mostly of  quantum origin). This problem is reminiscent of the interpretation of empty waves in the PWI where their peculiar energetic properties are often seen as a difficulty.\\
\indent Moreover  we now see that the problem disappears in our theory: The energy losses associated with radiated waves  (i.e., Fig.~\ref{image5}(b)) are compensated by  the energy gain  associated with the advanced waves converging on the particle (i.e., Fig.~\ref{image5}(c)).   From the point of view of usual causality   this looks conspiratorial or superdeterministic.  A `de Broglie-Bohm demon' having access to this flow of energy and perceiving the time going from past to future would see a converging flow of energy coming from the remote space arriving precisely at the good time in a coherent way on the particle.  Furthermore,  a retarded wave is also emitted by the particle and the sum off both  waves gives the soliton field discussed in Secs.~\ref{sec2} and \ref{sec3} imposing the guidance formula.  As we showed in Sec.~\ref{sec2} it is possible to build a stationary soliton field in the local rest frame.  When merging this near-field with the far-field of Sec.~\ref{sec3} the time-symmetric structure is thus required for consistency.  Therefore, in our model the non-linearity of the wave equation  and the existence of stable stationary solitons involves a time-symmetric causality. This in turn allows us to preserve energy conservation (more on this is derived in Appendix D) and reproduce the predictions of the PWI, i.e., of quantum mechanics (in the regimes considered here).\\
\indent The theory discussed in this work focused on the single particle/soliton problem and we showed that a time-symmetric $u-$field is required. This time-symmetric causality is clearly of great importance concerning the problem of entanglement between several  particles. As it is well-known in de Broglie Bohm mechanics~\cite{Hiley} non-locality is offered as an explanation for justifying violations of Bell's inequality.  However, the present  theory is definitely local  and its quantitative predictions should therefore apriori differ from the standard PWI. It is here  that the time-symmetry of the model comes to the rescue.\\
\indent To see how it works, we consider the many-body generalization to an ensemble of $N$ indistinguishable (bosonic) particles of the LKG equation for a single particle developed in Sec.~\ref{sec2}.   We have the wavefunction $\Psi_N(x_1,...x_i,...x_N)$  solution of the set of $N$ coupled equations $D_i^2\Psi_N=-\omega_0^2\Psi_N$ equivalent to the set of hydrodynamic equations\footnote{We have $D_i=\partial_i+eA(x_i)$, $\partial_i:=\frac{\partial}{\partial x_i}$ and the polar form $\Psi_N(x_1,...x_i,...x_N)=a_N(x_1,...x_i,...x_N)e^{iS_N(x_1,...x_i,...x_N)}$.}:                      
\begin{subequations}
\begin{eqnarray}
(\partial_i S_N(X)+eA(x_i))^2=\omega_0^2+\frac{\Box_i a_N(X)}{a_N(X)}:=\mathcal{M}^2_{\Psi_N,i}(X) \label{2N}\\
\partial_i[a_N^2(X)(\partial_i S_N(X)+eA(x_i))]=\partial_i[a_N^2(X)\mathcal{M}_{\Psi_N,i}(X)v_{\Psi_N,i}(X)]=0,\label{2Nb}
\end{eqnarray}
\end{subequations} where we introduced the notation $X:=[x_1,...,x_N]$ and the velocity $v_{\Psi_N,i}(X):=-\frac{\partial_i S_N(X)+eA(x_i)}{\mathcal{M}_{\Psi_N,i}(X)}$ (we also assume $\mathcal{M}^2_{\Psi_N,i}(X) >0$). In the PWI we define the velocity of the $N$ particles through the guidance relations
\begin{eqnarray}
\frac{dz_i(\lambda)}{d\tau_i}=\frac{\dot{z}_i(\lambda)}{\sqrt{\dot{z}_i(\lambda)\dot{z}_i(\lambda)}}=v_{\Psi_N,i}(Z(\lambda))\label{manybohm}
\end{eqnarray}   with $Z(\lambda):=[z_1(\lambda),...,z_N(\lambda)]$, $\dot{z}_i(\lambda):=\frac{dz_i(\lambda)}{d\lambda}$ and $d\tau_i=\sqrt{\dot{z}_i(\lambda)\dot{z}_i(\lambda)}d\lambda$ is a proper time element along the trajectory of the $i^{th}$ particle. We stress that in this description we require a  parameter $\lambda$ to synchronize the $N$ particles. Usually this is done by involving a preferred foliation $\mathcal{F}$ of Minkowski's space-time with space-like hyperplanes. Choosing $\mathcal{F}$ in general particularizes a set of entangled trajectories  defining an ensemble $M(\mathcal{F})$. Moreover, since the choice of $\mathcal{F}$ is arbitrary the  PWI admits an infinite number of possible  paths-ensemble  $M(\mathcal{F})$. Clearly, there is an apparent tension with relativity since the ensemble of trajectories  is not unique and depends on a foliation sometimes identified with a kind of `Bohmian-Aether' (for a discussion on this issue see \cite{Drezet2019}). In the context of our relativistic local theory  for a $u-$field we don't here give any ontological content to the particular   foliation used $\mathcal{F}_0$  to specify the particle trajectories. Instead, it is the (infinite) ensemble of all the  $M(\mathcal{F})$ that exhausts the set of possibilities; and  the choice  $\mathcal{F}_0$  used in our Universe (or in the part of our Universe accessible to us and entangled with us) is associated with a particular choice on initial conditions  (perhaps related to  cosmological constraints~\cite{Drezet2019}). We stress that the dynamics Eq.~\ref{manybohm} is not in general `statistically transparent', i.e., that it cannot always reproduce Born's rule and the statistical predictions of quantum mechanics (more on this will published in a subsequent article).  Moreover, in the non-relativistic regime  or in finite asymptotic regions of space-time, i.e., before or after scattering or interactions with an external field, we can justify Born's rule and recover statistical transparency.\\
\indent In the present local theory for the $u-$field the $N$ solitons move in the same 4D space time and not in the abstract configuration space.   The far-field for the $N$ entangled soliton is written in analogy with Eq.~\ref{debroglieSing3AS} as:
\begin{eqnarray}
 u(x)=\sum_i u_i(x) =\sum_i g(0)\int_{(C_i)}K_{sym}(x,z_i(\lambda))\frac{e^{iS_N(Z(\lambda))}}{\sqrt{\alpha_i(\lambda)}}d\tau_i \label{debroglieSing3ASMany}
\end{eqnarray} with $\alpha_i(\lambda)=\sqrt{\frac{\mathcal{M}_{\Psi_N,i}(Z(\lambda))}{\mathcal{M}_{\Psi,i}(Z(0))}}$. This field obeys the following local equation: 
\begin{eqnarray}
D^2 u(x)=\sum_i g(0)\int_{(C_i)}\delta^{4}(x-z_i(\lambda))\frac{e^{iS_N(Z(\lambda))}}{\sqrt{\alpha_i(\lambda)}}d\tau_i \label{debroglieSingtriMany}
\end{eqnarray} that is defined in the 4D Minkowski spacetime not in the configuration space. This wave equation and dynamics is local  but the $N$ singularities moving along the $N$ trajectories $C_i$ are clearly entangled through the phase $S_N(Z_i(\lambda))$ and the masses $\mathcal{M}_{\Psi_N,i}(Z(\lambda))$ defined in the PWI of Eqs.~\ref{2N},\ref{2Nb}. The $N$ trajectories are synchronized using the parameter $\lambda$ and a specific foliation $\mathcal{F}$.\\        
%%%%%%%% 
%%%%%%%%%%%%%%%%
\begin{figure}[h]
\centering
\includegraphics[width=8 cm]{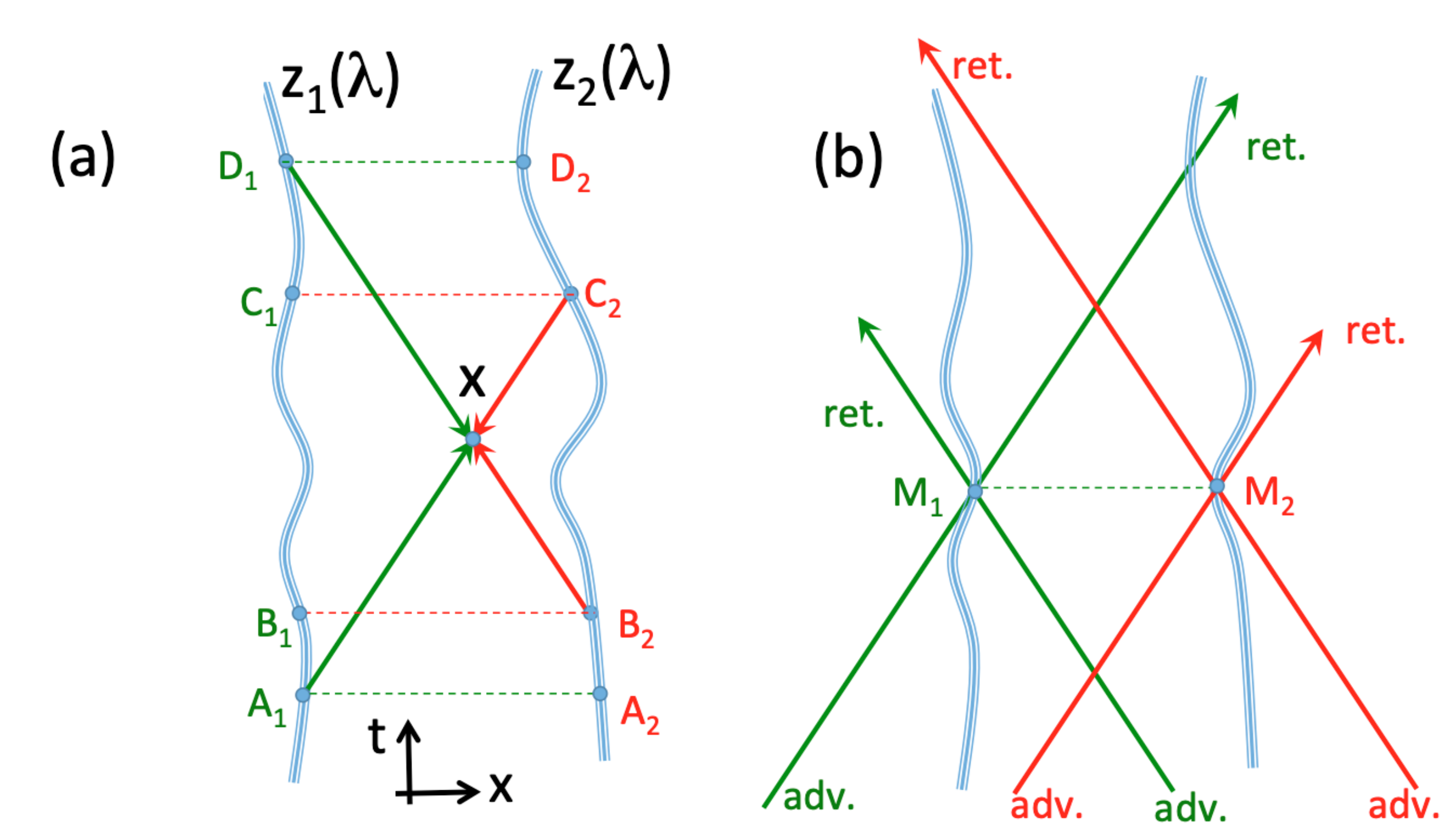}
\caption{(a) Sketch of $u-$field  created at space-time point $x$ by two entangled particles 1 and 2 with trajectories $z_1(\lambda),z_2(\lambda)$. The retarded field emitted by particle 1 at point $A_1$  (forward in time  green arrow) depends on the knowledge of the position of the particle 2 located at $A_2$. The same is true for the advanced field created at point $D_1$ by particle 1  (backward in time green arrow) and depending on the position of particle 2 at $D_2$.   The situation is the similar for the field created by particle 2 (red arrows) at $B_2$ and $C_2$ synchronized with the particle 1  at $B_1$ and $C_1$.  (b) The particle 1 at $M_1$ and 2 at $M_2$ are excited by advanced waves coming from the past and in turn emit retarded waves going into the future. The synchronization and correlation of the two particles are related to the time-symmetric causal connection with information located in the remote past and future.    }    \label{image6}
\end{figure}
%%%%%%%%% 
\indent Consider for example a pair of particles 1 and 2 with entangled trajectories $z_1(\lambda),z_2(\lambda)$ defined with the PWI guidance formula Eq.~\ref{manybohm}.  The $u-$field at point $x$ is the sum of the  contributions $u_1(x)$ and $u_2(x)$ generated by the particle $1$  and $2$ respectively. In the example of Fig.~\ref{image6}(a) (i.e., in absence of external $A-$field) $u_1(x)$ splits into  a retarded contribution emitted from the particle $1$ when it was at $A_1=[t_{A_1},\mathbf{x}_{A_1}]$ and an advanced contribution emitted  from the same particle when it was at $D_1=[t_{D_1},\mathbf{x}_{D_1}]$. Moreover, when the particle $1$ is at $A_1$ ($D_1$) the second particle is at point $A_2$ ($D_2$) as determined by the preferred foliation $\mathcal{F}$ (here we consider the set of hyperplanes $t=const.$ for such a foliation). We can use Eq.~\ref{Lienard} to evaluate $u_1$:
\begin{eqnarray}
 u_1(x)=\frac{g(0)}{2}\left([\frac{e^{iS(z_1(\lambda),z_2(\lambda))}}{4\pi\sqrt{\alpha_1(\lambda)}\rho_1(\lambda)}]_{\lambda=t_{A_1}}+[\frac{e^{iS(z_1(\lambda),z_2(\lambda))}}{4\pi\sqrt{\alpha_1(\lambda)}\rho_1(\lambda)}]_{\lambda=t_{D_1}}\right)\label{Lienard2}
 \end{eqnarray} with $\rho_1(\lambda)=|(x-z_1(\lambda))\cdot \frac{\dot{z}_i(\lambda)}{\sqrt{\dot{z}_1(\lambda)\dot{z}_1(\lambda)}}|$ and $\alpha_1(\lambda)=\sqrt{\frac{\mathcal{M}_{\Psi,1}(z_1(\lambda),z_2(\lambda))}{\mathcal{M}_{\Psi,1}(z_1(0),z_2(0)))}}$ for $\lambda=t_{A_1}$ or  $\lambda=t_{D_1}$. Of course, the field $u_2(x)$ generated by particle 2  is obtained with the same method and (as shown in Fig.~\ref{image6}(a)) it will involves  points   $B_2$ (of retarded emission by particle 2), $C_2$ (of advanced emission by particle 2)  and the correlated positions  of particle 1 at points $B_1$ and $C_1$. The total field is $u_1(x)+u_2(x)$.\\
\indent  More generally, the $u-$field obtained with Eq.~\ref{debroglieSing3ASMany} shows a mixture of local and non-local properties.  The local part is clearly the presence of the propagator $K_{sym}(x,z_i(\lambda))$  associated with the field equation Eq.~\ref{debroglieSingtriMany}. The non-local elements are associated with the  correlated phase $S_N(Z(\lambda))$ and masses $\mathcal{M}_{\Psi_N,i}(Z(\lambda))$ reminiscent of the PWI using the preferred foliation $\mathcal{F}$. In this approach the $u-$field propagates locally in the 4D spacetime but the singularities  are non-locally correlated. Moreover, in our theory  this is the local $u-$field  that is more fundamental and not the non-local (contingent) $\Psi-$wave. There is an  other way to watch this.  Indeed, the theory is time-symmetric and as illustrated in Fig.~\ref{image6}(b) each particle is fed by an advanced $u-$wave coming from the remote past whereas it emits a retarded wave propagating into the future. This ensures energy/momentum conservation for the $u-$field and also provides an explanation for the synchronization  and entanglement of the particles. A `de Broglie-Bohm demon'  watching the problem from past to future (i.e., as a Cauchy problem) will explain the `spooky' correlations between the particles as a superdeterministic consequence of the field preparation in the remote past. Don't forget: From Green's theorem  the total field reads $u(x)=\frac{u_{ret}(x)+u_{adv}(x)}{2}=u_{ret}(x)+u_0(x)$ where $u_0(x)=\frac{u_{adv}(x)-u_{ret}(x)}{2}$ is a solution of the homogeneous wave equation that can be interpreted as a free field exciting the particles in a conspiratorial looking way. But here the theory is time-symmetric as required for the solitons stability: Therefore the conspiracy is actually explained.\\
\indent Furthermore, don't also forget that Eq.~\ref{debroglieSingtriMany} involving Dirac distributions, and the entangled trajectories $C_i$ is just an effective wave equation for the far-field of the solitons.  Fundamentally   the only wave equation is $D^2u(x)=-N(u^\ast (x)u(x))u(x)$ i.e., Eq.~\ref{1b} that is non-linear but completely local. The separation $u(x)\simeq u_i(x)$ is just a very good approximation if the solitons are not  too close from each other.  If we approach the $i^{th}$ soliton we get $u(x)\simeq u_i(x)$ which is (an approximate) solution of  Eq.~\ref{1b}.  Now, we can apply the phase-harmony condition developed in Sec.~\ref{sec2} for a single soliton and here we get:
\begin{eqnarray}
\varphi_i(x) \simeq S_N(Z(\lambda)) -eA(z_i(\lambda))\xi_i+B_i(z_i(\lambda))\frac{\xi_i^2}{2}+O(\xi_i^3)\label{phaseharmonymany}
\end{eqnarray} where $\xi_i=x-z_i(\lambda)$ is defined in the local proper hyperplane $\Sigma_i(\lambda)$ normal to the velocity $\frac{dz_i(\lambda)}{d\tau_i}$ given by Eq.~\ref{manybohm}. $B_i(z_i(\lambda))$ is a collective coordinate defined as in Sec.~\ref{sec2}. All the deductions and theorems discussed in Sec.~\ref{sec2} are still valid (don't forget we use $d\tau_i=\sqrt{\dot{z}_i(\lambda)\dot{z}_i(\lambda)}d\lambda$). This allows us to build the near-field of each soliton looking  like the monopolar field (see Eq.~\ref{LanesolutionREscaled}): \begin{eqnarray}
F_{i,\lambda}(r_i)=\frac{\sqrt{\alpha_i(\lambda)}g(0)}{4\pi}\frac{1}{\sqrt{\alpha_i(\lambda)^2r_i^2+r_0(0)^2}}
\label{LanesolutionREscaledmany}
\end{eqnarray} with $r_i=\sqrt{(-\xi_i^2)}$ in  $\Sigma_i(\lambda)$.\\ 
\indent What is of course remarkable in Eq.~\ref{phaseharmonymany}, is the presence  of the nonlocal phase $S_N(Z(\lambda))$ associated with $\Psi_N(Z(\lambda))$. Even if $u(x)\simeq u_i(x)=f_i(x)e^{i\varphi_i(x)}$ is a local field solution of Eq.~\ref{1b} nothing prohibits us to use the non-local phase $S_N(Z(\lambda))$ obtained from $\Psi_N(X)$ and evolving in the configuration space. No violation of the conservation laws for the $u-$field will appear by doing this choice which is therefore completely legitimate.   In that sense there is a gentle agreement between nonlocal and local effects in our theory.   Non-locality is only an effective property allowed by the nonlinear and time-symmetric  $u-$field  used in our approach. This clearly defines a new paradigm where a local theory is able to reproduce the nonlocal properties of the PWI.\\
\indent More generally, we found remarkable that in the new paradigm all the elements are strongly connected and related together to make the theory working fine.    Nonlinearity  and time-symmetry are required for the stability of our solitons  and at the same time justify the existence of a guidance formula needed for deriving the PWI.  The time-symmetry  modifies the usual causality from past to future and allows for emerging  and effective nonlocal features (i.e., in agreement with Bell's theorem). In that sense nonlocality emerges from local physics in a consistent way.\\      
%%%%%%%%%%%%%%%%
\begin{figure}[h]
\centering
\includegraphics[width=6 cm]{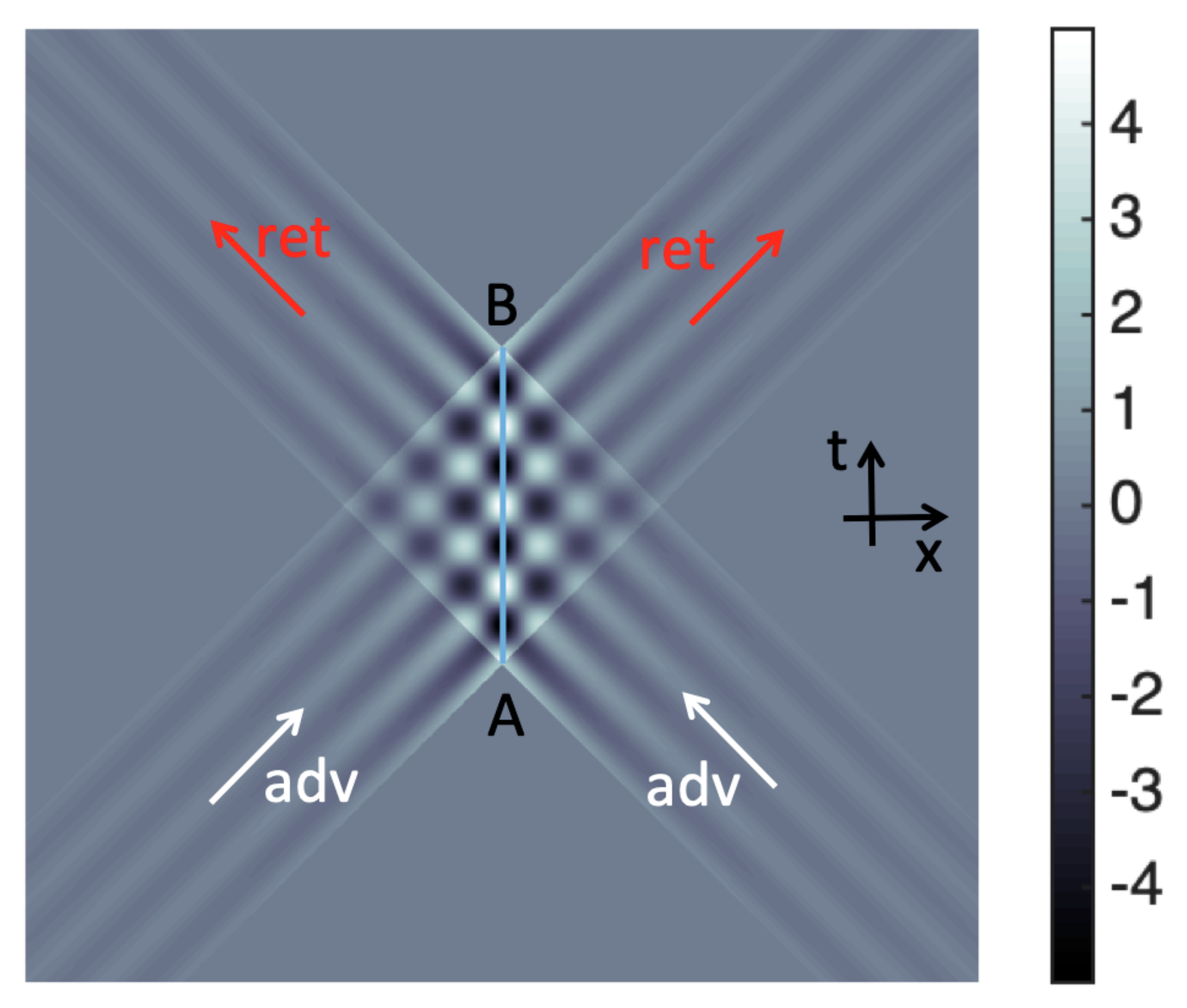}
\caption{2D map of $\textrm{Re}[u(x)]$ for a time symmetric monopole created at $A$ and destroyed at $B$.   Advanced waves  (white arrows) coming from the remote past converge to the particle at rest (vertical blue line). Retarded waves are emitted in to the future direction (red arrows). The diamond shape results from the interference between both waves and is associated with the stationary particle soliton.  }    \label{image7}
\end{figure}
%%%%%%%%%        
\indent An other remarkable feature of our soliton model is that it circumvents the conclusions obtained in \cite{Submitted} with Ehrenfest's theorem for a strongly localized soliton.   Here our soliton associated with a monopole  $\sim 1/r$ at large distance is not sufficiently localized to impose a classical-like dynamics.  The deformation of the soliton obtained in our model allows him to follow a de Broglie-Bohm dynamics. Again, this is strongly related to the other features of the model discussed above.   One interesting aspect of this weak localization must be emphasized.  Indeed, consider a soliton at rest in free space with the monopolar field given by Eq.~\ref{interpolate1} reducing to de Broglie solution Eq.~\ref{debroglie1}  in the far-field. The full energy of the soliton  is given by $E=\int d^3\mathbf{x}[U(f^2)-f^2N(f^2)+2\omega_0^2f^2+\boldsymbol{\nabla}(f\boldsymbol{\nabla}f)]$ that can approximately written as
\begin{eqnarray}
E\simeq E_s+\frac{g^2}{4\pi}[\omega_0^2 R-\frac{\textrm{cos}^2(\omega_0 R)}{R}]\label{infi}
\end{eqnarray}   where $E_s=\int d^3\mathbf{x}[U(f^2)-f^2N(f^2)]$ has been evaluated using the near-field\footnote{The error is small in the integration since $U(f^2)-f^2N(f^2)\sim 1/r^6$ at large distances. }, i.e., $E_s=\frac{g^2}{32r_0}$ (see Appendix A). The integration in Eq.~\ref{infi} has been pushed until a large radius $R$. In the limit $R\rightarrow +\infty$ $E$ contains a diverging contribution growing linearly with $R$ the other term goes to zero as $1/R$. A diverging energy seems at first pathological. However, note that if the particle has a finite life-time $T$ the radius $R$ cannot grow indefinitely. As illustrated in Fig.~\ref{image7} if the particle appears at $A$ (time $t=0$) and disappears at $B$ (time $t=T$) the $u$-field must have a diamond like shaped structure where advanced waves coming from the past direction interfere with retarded waves emitted to the future direction and create the stationary field of Eq.~\ref{debroglie1}. The diamond structure of Fig.~\ref{image7} is  built between the light cones coming from past and future. Integrating the total energy  at times $t<0$ or $t>T$ gives the approximate value $E\simeq \frac{g^2}{4\pi}\omega_0^2 \frac{T}{2}$ associated with advanced or retarded waves.   During the time interval $0\leq t\leq T$ we obtain $E=E_s+\frac{g^2}{4\pi}\omega_0^2 \frac{T}{2}$ the difference is attributed to the local formation of the particle at $A$ requiring an additional energy  $E_s$ coming from the environment at $A$. This energy is returning to the environment at $B$ when the particle disappears\footnote{This description made in the regime $\omega_0T\gg1$ is of course an approximation that neglects the transient effects associated with the discontinuities at $A$ and $B$ contributing to the energy balance. }.\\
\indent We can naturally speculate on the scale $R$ at which the far-field energy $E_{ff}=\frac{g^2}{4\pi}\omega_0^2 R$ in Eq.~\ref{infi} becomes comparable to $E_s$. The ratio $\frac{E_{ff}}{Es}=32\pi\frac{r_0}{\lambda_0}\frac{R}{\lambda_0}$ depends on the size of the particle $r_0$ and the Compton wavelength $\lambda_0=2\pi/\omega_0$. In absence of a more precise theory fixing the value of $r_0$ the ratio is let undetermined. Moreover, since $r_0$ is supposed to be very small an effect should only be observed at astrophysical or cosmological scales. For instance consider a proton  with $\lambda_0\sim 10^{-15}$ m  and suppose  $R_G=10^{16}$ m a typical size for a galaxy. If $r_0$ is of the order of $r_0\sim 10^{-48}$ m (i.e., much smaller than the Planck length $r_P=10^{-35}$ m) we obtain $\frac{E_{ff}}{Es}\sim 1$. Interestingly, $R_G$ is the scale at which dark-matter is usually involved in order to explain the rotation curve anomaly of stars in spiral galaxy. As it is known,  the density of dark-matter needed to explain the constant value of the star velocity $v_\infty$ at large distance of the galaxy core  is typically growing as $\rho_{DM}(R)\propto 1/R^2$ and the mass as $M_{DM}(R)\propto R$ for $R\sim R_G$. This is typically what we obtain in our soliton model of quantum particles where the particle-core with energy $E_s$ is surrounded by a halo of energy (mass) growing as $E_{ff}\propto R$. With a value of  $r_0\sim 10^{-48}$  m our model could thus potentially explains the anomaly in the rotation curves and interpret dark-matter as the far-field gravitational contribution of the particle masses to the dynamic of galaxies. Of course this is very speculative, and  in the end it is not yet very clear what is the status of the particle energy $E$ in our theory. We point out that the conserved norm $\mathcal{Q}=\int d^3\mathbf{x}2\omega_0^2f^2$ associated with the current conservation Eq.~\ref{2d} can be computed for the same example leading to Eq.~\ref{infi}. We get 
\begin{eqnarray}
\mathcal{Q}(R)\simeq \frac{g^2}{4\pi}[\omega_0 R+\frac{\textrm{sin}(2\omega_0 R)}{2}].\label{infiB}
\end{eqnarray}    In the limit $R\rightarrow +\infty$ we obtain $\frac{E}{\mathcal{Q}}\simeq\omega_0[1+\frac{\lambda_0^2}{32\pi r_0 R}]\rightarrow\omega_0$, i.e., $\hbar\omega_0$ the quantum energy formula. So perhaps it is the ratio $\frac{E}{\mathcal{Q}}$ that should be identified with the physical energy of the soliton. This could be important when considering coupling with the gravitational field where a definition of mass must be included.\\
\indent To conclude this work, it is important to mention that several important questions are left open and unanswered.   For example, in our model we ignored the self-electromagnetic field generated by  the soliton.  This can be a good approximation near the particle core but from Eq.~\ref{infiB} we see that the electric charge $q(R)$ contained in a sphere of radius $R$ centered on the soliton is given by $q(R)=e\mathcal{Q}(R)\simeq\frac{eg^2}{4\pi}\omega_0 R$ which diverges linearly. From Gauss's theorem this implies a radial electric field $\mathbf{E}(R)=\frac{eg^2}{(4\pi)^2}\frac{\omega_0}{R}\mathbf{\hat{R}}$ different from the standard Coulomb's field.  This problem could be  perhaps solved by renormalizing the electric  charge or by imposing the constraint $|q(R_U)|\ll |e|$ for a large radius $R_U\sim 10^{26} $ m  (size of the observable Universe).  This implies  $\frac{g^2}{2}\frac{R_U}{\lambda_0}\ll 1$ and for a proton we need $g^2\ll 10^{-41}$.  If this is true we could neglect the electromagnetic coupling between solitons\footnote{Of course the problem is absent if we limit the present model to neutral solitons  with $e=0$.}.  New ideas should be thus inserted in the model to develop electromagnetic interactions between solitons perhaps  mediated with  localized solitons associated with photons.   We also mention that the NLKG equation used here has some pathological features associated with the tachyonic sector alluded to briefly in Sec.~\ref{sec2}.   We restricted the analysis made in this work to the case of solitons with $\mathcal{M}_\Psi(z)^2>0$ but the tachyonic sector $\mathcal{M}_\Psi(z)^2<0$ was rejected as unphysical. Perhaps this could be avoided if the model is modified to incorporate the idea of `fusion' developed by de Broglie  where a spin zero particle is understood as a composite object made of two solitons with spins $1/2$.  In the very end,similar approaches can  certainly be  developed for particles with integer spins like photons or gravitons, or with Dirac spinors for generating solitonic fermions with spin $1/2$ (this will be discussed in subsequent articles).           
 %%%%%%%%%%%%%%%%%%%%%%%%%%%%%%%%%%%%%%%%%%%   Appendix    %%%%%%%%%%%%%%%%%%%%%%%%%%%%%%%%%%%%%%%%%%%%%%%%%%%%%%%%%%%%%%%%%%4%%%%%%%%%%%%%%%%%%%%%%%%%%%%%%%%%%%%%%%%
%%%%%%%%%%%%%%%%%%%%%
\section*{Appendix A}          
\label{appd} 
\indent Using Eqs.~\ref{Lane} and \ref{Lane2} we define the integral 
\begin{eqnarray}
	G=-\int d^3\textbf{x}N(f^2)f=\frac{3g}{2}\int_0^{+\infty}d\eta\frac{\sqrt{\eta}}{(\eta+1)^{\frac{5}{2}}}
\end{eqnarray} where $\eta=r^2/r_0^2$. By definition this is related to  the beta Euler function $B(\frac{3}{2},1)$ by 
	$G=\frac{3g}{2}B(\frac{3}{2},1)=\frac{3g}{2}\frac{\Gamma(\frac{3}{2})\Gamma(1)}{\Gamma(\frac{5}{2})}$ where $\Gamma(z)$ is Euler's Gamma function. We have finally 
\begin{eqnarray}
	G=\frac{3g}{2}\frac{\frac{\sqrt{\pi}}{2}}{\frac{3}{2}\frac{\sqrt{\pi}}{2}}=g.
\end{eqnarray}
\indent The static energy $E_s$ associated with the soliton given by Eq.~\ref{Lane2}
is by definition $E_s=\int d^3\textbf{x}[(\boldsymbol{\nabla}f)^2+U(f^2)]$.
 Inserting Eq.~\ref{Lane} leads after integration by part to
 \begin{eqnarray}
	E_s=\int d^3\textbf{x}[U(f^2)-N(f^2)f^2].\label{estatic}
\end{eqnarray} Using Eq.~\ref{Lane} and Eq.\ref{Lane0} we get
\begin{eqnarray}
	E_s=\frac{g^2}{4\pi r_0}\int_0^{+\infty}d\eta\frac{\sqrt{\eta}}{(\eta+1)^{3}}
=\frac{g^2}{4\pi a}B(\frac{3}{2},\frac{3}{2})
\end{eqnarray} which finally yields
\begin{eqnarray}
	E_s=\frac{g^2}{4\pi r_0}\frac{\Gamma(\frac{3}{2})\Gamma(\frac{3}{2})}{\Gamma(3)}
	=\frac{g^2}{4\pi r_0}\frac{(\frac{\sqrt{\pi}}{2})^2}{2}=\frac{g^2}{32r_0}.
\end{eqnarray} 
 %%%%%%%%%%%%%%%%%%%%%%%%%%%%%%%%%%%%%%%%%%%%%%%%%%%%%%%
\section*{Appendix B}          
\label{appDerrick}
\indent For a static soliton $f(\mathbf{x})\in \mathbb{R}$ solution of the equation 
\begin{equation}\boldsymbol{\nabla}^2f=N(f^2)f\label{appDerrick1}\end{equation} 
in the 3D space we can define the static energy $E_s=\int d^3\textbf{x}[(\boldsymbol{\nabla} f)^2+U(f^2)]$.   $E_s$ can be used to establish a variational principle $\delta E_s=0$ for recovering the field equation  $\boldsymbol{\nabla}^2f=N(f^2)f$. In \cite{Hobart,Derrick} the authors consider the stretching or dilation transformation $f(\mathbf{x})\rightarrow f(\alpha\mathbf{x})$ with $\alpha$ a positive real number. Here we instead consider the more general transformation 
\begin{equation}
f(\mathbf{x})\rightarrow \tilde{f}(\mathbf{x})=\beta f(\alpha\mathbf{x})\label{appDerrick2}
\end{equation}
 with $\beta\in \mathbb{R}$.\\
\indent Under this transformation we check that the new function $\tilde{f}(\mathbf{x})$ obeys Eq.~\ref{appDerrick1} iff we have $\alpha^2N(\tilde{f}^2/\beta^2)=N(\tilde{f}^2)$. Here we consider specifically the general Lane-Emden nonlinearity $N_p(x)=-\gamma x^p$ with $\gamma, p\in \mathbb{R}$  (the case $\gamma>0$, $p=2$ is the one considered in this article). Within this family of nonlinearity functions we obtain the constraint \begin{eqnarray}
\beta=\alpha^{1/p}\label{appDerrick2b}
\end{eqnarray} which reduces to $\beta=\sqrt{\alpha}$ used in the main text for $p=2$.
Moreover, by using Eq.~\ref{appDerrick2} and $N_p$ the static energy $E_s$ for $\tilde{f}$ becomes a function of $\alpha$ reading
 \begin{subequations}
 \begin{eqnarray}
 E_s(\alpha)=\frac{\beta^2 }{\alpha}I_k-\frac{\beta^{2(p+1)} }{\alpha^3}I_p \label{appDerrick3a}\\
 =\alpha^{2/p-1}(I_k-I_p)\label{appDerrick3b}
\end{eqnarray}
\end{subequations}   with $I_p=\frac{\gamma}{p+1}\int d^3\textbf{x}(f(\textbf{x}))^{2(p+1)}$ and $I_k=\int d^3\textbf{x}(\boldsymbol{\nabla}f(\textbf{x}))^2$. Importantly, in passing from Eq.~\ref{appDerrick3a} to \ref{appDerrick3b} we used the constraint  Eq.~\ref{appDerrick2b}.\\
\indent We now consider a first order variation $\delta E_s=E_s(\alpha)-E_s(1)$ with $\alpha=1+\epsilon$ and $\epsilon\rightarrow 0$.  In \cite{Hobart,Derrick} the authors imposed $\beta=1$ and if we use Eq.~\ref{appDerrick3a} the variational condition $\delta E_s=0$ implies
 \begin{eqnarray}
\left.\frac{dE_s}{d\alpha}\right\vert_{\alpha=1}=-I_k+3I_p=0.\label{appDerrick5}
\end{eqnarray} Therefore, we deduce $3I_p=I_k$ which is non negative by definition of $I_k$ and implies $\gamma>0$.    
Similarly, we can define a second order variation $\delta^2 E_s$ and we obtain
\begin{eqnarray}
\left.\frac{d^2E_s}{d\alpha^2}\right|_{\alpha=1}=2I_k-12 I_p=-6I_p<0.\label{appDerrick6}
\end{eqnarray} This implies unstability of the soliton. However, a physical transformation for this soliton must rely on the constraint Eq.~\ref{appDerrick2b} in order to fulfill Eq.~\ref{appDerrick3b}. Therefore instead of Eq.~\ref{appDerrick5} we must have:
 \begin{eqnarray}
 \left.\frac{dE_s}{d\alpha}\right|_{\alpha=1}=(2/p-1)(I_k-I_p)=0.\label{appDerrick5b}
\end{eqnarray} Moreover, we have $I_k-I_p=E_s(1)$ and $I_k=(p+1)I_p$ (as it can be checked after integration by parts of $I_k$ and neglecting a surface integral term)  and we thus get $(2/p-1)pI_p=0$ which imposes the value $p=2$ (this result was obtained in \cite{Rosen1966}). Observe that if we insert the formula $I_k=(p+1)I_p$ in Eq.~\ref{appDerrick5} we obtain $-(p+1)I_p+3I_p=0$ and therefore we again deduce the condition $p=2$ which is thus imposed by either Eq.~\ref{appDerrick5} or Eq.~\ref{appDerrick5b}. This result assumes that the field decays fast enough (i.e., at least as $f\sim 1/r^m$ with $m>1/2$ for $r$ large\footnote{Note that in order to have $I_p< \infty$ we must have  $m>\frac{3}{2(p+1)}$ so that   globally $m> max[\frac{1}{2},\frac{3}{2(p+1)}]$ \cite{Rosen1966}.}) in order to neglect the surface integral term in $I_k$.  Furthermore, with Eq.~\ref{appDerrick3b} we obtain 
\begin{eqnarray}
\left.\frac{d^2E_s}{d\alpha^2}\right|_{\alpha=1}=(2/p-1)(2/p-2)(I_k-I_p).\label{appDerrick6b}
\end{eqnarray} replacing  Eq.~\ref{appDerrick6}.  Clearly, from Eq.~\ref{appDerrick5b} we deduce $\left.\frac{d^2E_s}{d\alpha^2}\right|_{\alpha=1}=0$ which means that the soliton is not anymore unstable: it is metastable. This result evades the conclusions of the Hobart-Derrick theorem which was established without using the legitimate dilation transformation Eq.~\ref{appDerrick2b}. 
%%%%%%%%%%%    
\section*{Appendix C}          
\label{appg}
\begin{figure}[h]
\centering
\includegraphics[width=6 cm]{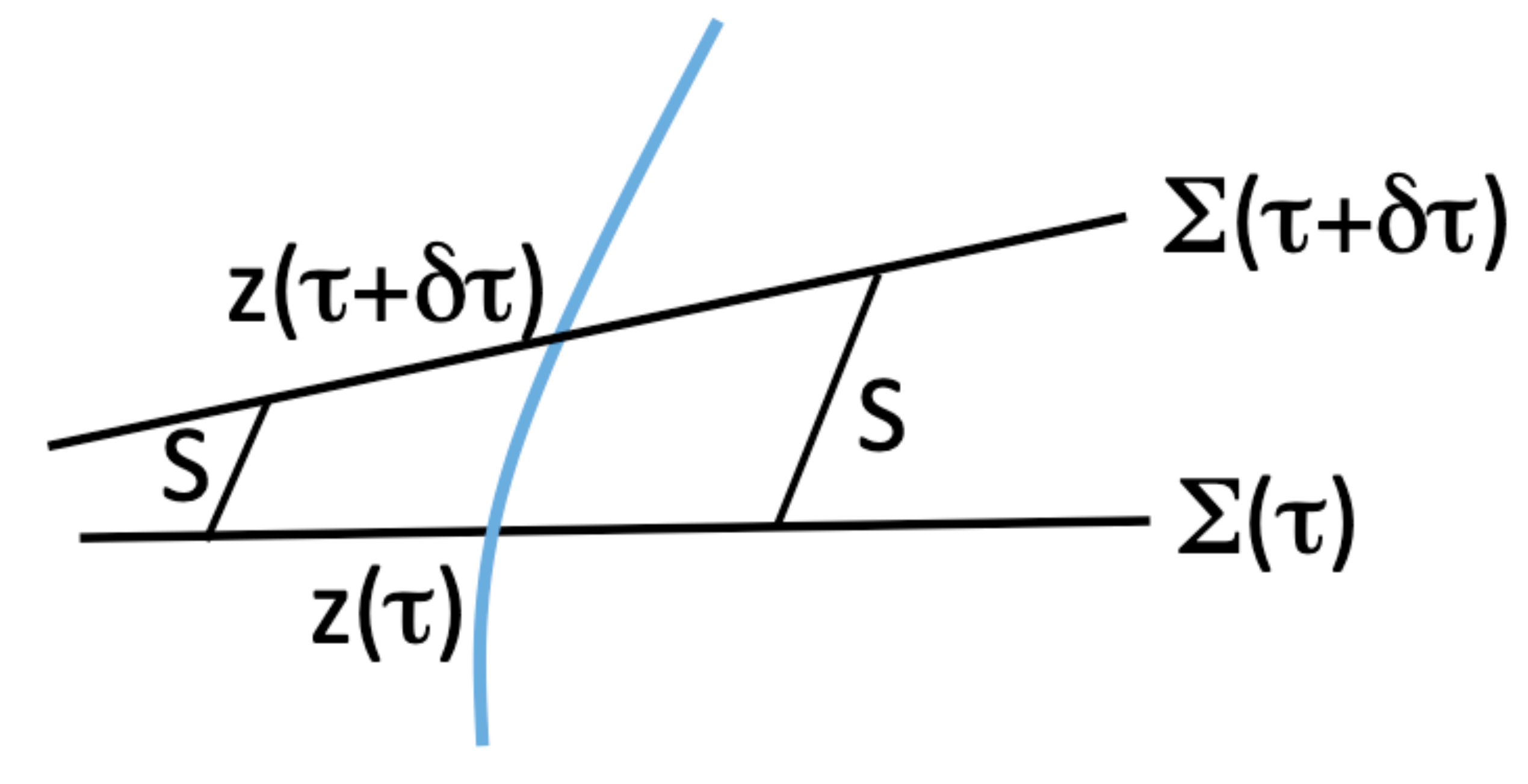}
\caption{Sketch of the 4D volume surrounding the particle trajectory $z(\tau)$ as discussed in the main text. 
}    \label{imageAPP}
\end{figure}

%%%%%%%%%%%%%%%%
\indent We start with the local current conservation $\partial_\mu(f^2(x)\mathcal{M}_u(x)v_u^\mu(x))=0$.
Consider now the 4D volume sketched in Fig.~\ref{imageAPP} which is bound by i) the two hyperplanes $\Sigma(\tau)$ and $\Sigma(\tau+\delta \tau)$ normal to   respectively $\dot{z}(\tau)$ and $\dot{z}(\tau+\delta\tau)$ (with $\delta \tau$ an infinitesimal delay time), and ii) the cylindrical hypersurface $S$ surrounding the particle trajectory.   This hypersurface $S$ is a 3D object which projects as a 2D closed surface surrounding the particle position $\mathbf{z}(\tau)$ in the hyperplane $\Sigma(\tau)$.
 The local height $\delta t$ of the cylinder is given by \cite{Submitted} $\delta t=\delta \tau(1-\xi\ddot{z}(\tau))$. A direct application of Gauss's theorem  applied to $\partial_\mu(f^2(x)\mathcal{M}_u(x)v_u^\mu(x))=0$ in this 4D volume in space-time leads to  
\begin{eqnarray}
\int_{\Sigma(\tau+\delta\tau)}f^2(x)\mathcal{M}_u(x)v^\mu_u(x)\cdot\dot{z}_\mu(\tau+\delta \tau)d^3\sigma\nonumber\\-\int_{\Sigma(\tau)}f^2(x)\mathcal{M}_u(x)v^\mu_u(x)\cdot\dot{z}_\mu(\tau)d^3\sigma\nonumber\\
=\oint_{S}d^2S_\mu v_u^\mu(x)(1-\xi\ddot{z}(\tau))\delta \tau\mathcal{M}_u(x)
\end{eqnarray}
 with $d^2S^\mu:= [0,d^2\mathbf{S}]$ a 4-vector associated with the local surface $d^2\mathbf{S}$ of the 2D surface surrounding $\mathbf{z}(\tau)$.   We have  
\begin{eqnarray}
\oint_{S}d^2S_\mu v_u^\mu(x)(1-\xi\ddot{z}(\tau))\delta \tau\mathcal{M}_u(x)\nonumber\\=-\oint_{S}d^2\mathbf{S}\cdot\mathbf{v}_u(x)[1+\boldsymbol{\xi}\cdot\ddot{\mathbf{z}}]\delta \tau\mathcal{M}_u(x).\label{g1}
\end{eqnarray}
Moreover, we have near the soliton center $\mathbf{v}_(x)\simeq O(\boldsymbol{\xi})$. Therefore, writing $d^2\mathbf{S}=\boldsymbol{\xi}^2d^2\Omega$ (where  $d^2\Omega$ is an elementary solid angle) the surface integral in Eq.~\ref{g1} is varying like   $O(\xi^3)$  which is neglected.   Similarly,  for  the scalar products of the velocities we have $v^\mu_u(x)\cdot\dot{z}_\mu(\tau+\delta \tau)\simeq 1$ and  $v^\mu_u(x)\cdot\dot{z}_\mu(\tau)\simeq 1$ and Eq.~\ref{g1} reduces to:
  \begin{eqnarray}
\frac{d}{d\tau}\int_{\Sigma(\tau)}f^2(x)\mathcal{M}_u(x)d^3\sigma\simeq 0
\end{eqnarray} which leads to 
\begin{eqnarray}
\int_{\Sigma(\tau)}f^2(x)\mathcal{M}_u(x)d^3\sigma=C
\end{eqnarray} where $C$ is a constant (assuming the volume  small).\\
\indent Now, writing $\mathcal{M}_u(x)\simeq \mathcal{M}_\Psi(z(\tau))$ and using Eq.~\ref{LanesolutionREscaled} we have $f(x)=F_\tau(r)=\sqrt{\alpha(\tau)}F_0(\alpha(\tau)r)$. After using the variable $\mathbf{w}=\alpha(\tau)\boldsymbol{\xi}$ we have 
\begin{eqnarray}
C=\mathcal{M}_\Psi(z(\tau))\frac{1}{\alpha^2(\tau)}\int F_0(w)d^3\boldsymbol{w}
\end{eqnarray}which directly leads to Eq.~\ref{inteF}. 
%%%%%%%%%%%    
\section*{Appendix D}          
\label{appenergy}
\indent Local energy-momentum conservation for the field of Eq.~\ref{1b} can be written in different equivalent ways. Here using the hydrodynamic formalism  we introduce a energy-momentum tensor $T^{\mu\nu}(x)=2f^2(x)\mathcal{M}_u^2(x)v_u^\mu(x)v_u^\nu(x)$ obeying the conservation law:
\begin{eqnarray}
\partial_\mu T^{\mu\nu}=2f^2\mathcal{M}_u\frac{d}{d\tau_u}(\mathcal{M}_u v_u^\nu)=2f^2\mathcal{M}_u[\partial^\nu(\mathcal{M}_u)+eF^{\nu\mu}{v_u}_\mu]    \label{conservationenergy}
\end{eqnarray}  with $\frac{d}{d\tau_u}:=v_u^\mu\partial_\mu$. We used the current conservation to obtain the first equality. This leads to $\frac{d}{d\tau_u}(\mathcal{M}_u v_u^\nu)=\partial^\nu(\mathcal{M}_u)+eF^{\nu\mu}{v_u}_\mu$ that can be obtained directly from Eq.~\ref{2c} and represents a quantum generalization of Newton's force formula for the $u-$field.\\
\indent A different way to write the energy-momentum conservation law is:
\begin{eqnarray}
\partial_\mu [T_0^{\mu\nu}+\eta^{\mu\nu}V(f^2)]=2ef^2\mathcal{M}_uF^{\nu\mu}{v_u}_\mu
\end{eqnarray} with $T_0^{\mu\nu}=T^{\mu\nu}+2\partial^\mu f\partial^\nu f-\eta^{\mu\nu}[(\partial f)^2+f^2\mathcal{M}_u^2]$. Finally, if we consider the full Maxwell's equations we have $\partial_\mu F^{\mu\nu}=2ef^2\mathcal{M}_u {v_u}^\nu$ and therefore if we write $T_{em}^{\mu\nu}$ the standard electromagnetic field  energy-momentum tensor we must have $\partial_\mu T_{em}^{\mu\nu}=-2ef^2\mathcal{M}_uF^{\nu\mu}{v_u}_\mu$. In the end we get: 
 \begin{eqnarray}
\partial_\mu [T_0^{\mu\nu}+\eta^{\mu\nu}V(f^2)+T_{em}^{\mu\nu}]=0.
\end{eqnarray}
In the Appendix D of \cite{Submitted} we applied Gauss's theorem to Eq.~\ref{conservationenergy} in a 4-D world tube surrounding the trajectory $z(\tau)$ of a soliton with  two ending  (3D) spacelike hyper-surfaces $\delta\Sigma_{A}$ and $\delta\Sigma_{B}$, and obtained:
\begin{eqnarray}
\int_{\delta\Sigma_{B}}2f^2(x)\mathcal{M}_u^2v_u^\nu-\int_{\delta\Sigma_{A}}2f^2(x)\mathcal{M}_u^2v_u^\nu\\
=\int_A^B d\tau\int_{\Sigma_0(\tau)}d^3\sigma_0 2f^2\mathcal{M}_u[\partial^\nu\mathcal{M}_u+eF^{\nu\mu}{v_{\nu}}_u].\label{tube}
\end{eqnarray}  In \cite{Submitted} we showed that for a strongly localized soliton like an undeformable Gausson this relation leads to a form of Ehrenfest's theorem where the quantum potential term cancels out because  the $u-$field amplitude decays exponentially far away from $z(\tau)$.   In the present work with a weakly localized soliton  with $f\sim 1/r$ at large distance from $z(\tau)$ we can not apply this result.  Moreover,  taking infinitely small cross-sections $\delta\Sigma_{\tau}$ for the tube and using the fact that near the trajectory $z(\tau)$ we have (see the footnote 2): $\mathcal{M}_u(x)\simeq \mathcal{M}_\Psi(z(\tau))+\boldsymbol{\xi}\cdot\boldsymbol{\nabla}\mathcal{M}_\Psi(z(\tau))+O(\xi^2)$, $\partial_\mu\mathcal{M}_u(z(\tau))=\partial_\mu\mathcal{M}_\Psi(z(\tau))$. This can be easily used to justify once more the dynamical law
$\frac{d}{d\tau_u}(\mathcal{M}_\psi(z) \dot{z}^\nu)=\partial^\nu(\mathcal{M}_\psi(z))+eF^{\nu\mu}\dot{z}_\mu$ associated with the PWI.
\section*{Competing Interest}
The Author declares no competing interest for this work. 

 \section*{Data Availability Statement}
Data Availability Statement: No Data associated in the manuscript. 
%%%%%%%%%%%%%%%%
%%%%%%%%%%%%%%%%%%%%%%%%%%%%%%%%%%%%%%


 
\end{document}